\documentclass[twocolumn,trackchanges]
{aastex631}
\usepackage{placeins}
\usepackage{float}
\usepackage{chngpage}

\usepackage{tikz}
\usetikzlibrary{arrows,calc,positioning}

\tikzstyle{intt}=[draw,text centered,minimum size=6em,text width=5.25cm,text height=0.34cm]
\tikzstyle{intl}=[draw,text centered,minimum size=2em,text width=2.75cm,text height=0.34cm]
\tikzstyle{int}=[draw,minimum size=2.5em,text centered,text width=3.5cm]
\tikzstyle{intg}=[draw,minimum size=3em,text centered,text width=6.cm]
\tikzstyle{sum}=[draw,shape=circle,inner sep=2pt,text centered,node distance=3.5cm]
\tikzstyle{summ}=[drawshape=circle,inner sep=4pt,text centered,node distance=3.cm]
\usetikzlibrary{shapes.geometric, arrows}
\tikzstyle{arrow} = [thick,->,>=stealth]

\usepackage{enumitem}

\usepackage{orcidlink}
\usepackage{multirow}

\usepackage[caption=false]{subfig}

\usepackage{amsmath}
\shorttitle{Electron Beam Aurorae}
\shortauthors{Zuckerman et al.}

\begin{document}

\title{Simulations of Electron Beam Interactions in Brown Dwarf Atmospheres}

\correspondingauthor{Zuckerman}
\email{anna\_zuckerman@alumni.brown.edu}

\newcommand{\CUB}{Department of Astrophysical and Planetary Sciences, University of Colorado Boulder, 2000 Colorado Ave, Boulder, CO 80309}
\newcommand{\UT}{Department of Astronomy, University of Texas at Austin, 2515 Speedway, Austin, TX 78712}
\newcommand{\LASP}{Laboratory for Atmospheric and Space Physics, 3665 Discovery Drive Boulder, CO 80303}

\author[0000-0002-2412-517X]{Anna Zuckerman} 
\altaffiliation{NSF Graduate Research Fellow}
\affiliation{\CUB}
\affiliation{\LASP}

\author[0000-0002-4489-0135]{J. Sebastian Pineda}
\affiliation{\CUB}
\affiliation{\LASP}  

\author[0000-0001-8932-368X]{David Brain}
\affiliation{\CUB}
\affiliation{\LASP}

\author[0000-0001-5864-9599]{James Mang}
\altaffiliation{NSF Graduate Research Fellow}
\affiliation{\UT}

\author[0000-0002-4404-0456]{Caroline V. Morley}
\affiliation{\UT}



\begin{abstract}

Over two decades ago, the first detection of electron cyclotron maser instability (ECMI) radio emission from a brown dwarf confirmed the presence of aurorally precipitating electrons on these objects. This detection established that brown dwarfs can exhibit magnetic activity that is planetary and auroral, rather than stellar in nature. This discovery motivated ongoing observational searches for the corresponding optical, ultraviolet (UV), and infrared (IR) auroral emission expected based on solar system analogs. The continuing nondetection of such auroral emission indicates important differences exist between auroral processes on brown dwarfs and solar system planets. In this work, we implement a Monte Carlo simulation of monoenergetic electron beams interacting with brown dwarf atmospheres, as a step towards understanding the physics of brown dwarf auroral emission. We detail the algorithm and underlying assumptions, and validate against previously published Jovian results \citep{Hiraki2008}. Our results agree well with literature, with some discrepancy from our updated interaction cross sections. We demonstrate the applicability of our simulation across the range of surface gravities and effective temperatures of radio-emitting brown dwarfs. We present an analytic parameterization of interaction rates based on our finding that atmospheric column density governs the interaction profiles. We apply this parameterization to calculate the total volumetric interaction rates and energy deposition rate for representative electron beam energy spectra enabling future predictions for spectra of aurorally emitting brown dwarfs. Simulations of high energy electron interactions with substellar hydrogen-dominated atmospheres will guide observational searches for multi-wavelength auroral features beyond the solar system.

\end{abstract}



\section{Introduction} \label{sec:intro}

Increasingly strong observational evidence indicates that brown dwarfs exhibit aurorae similar to those observed from solar system gas giants \citep{Hallinan_2015, Pineda2017, Kao2016}, but that the mechanisms behind the auroral emission must differ significantly \citep{Pineda2024, Saur2018}. Radio observations of brown dwarfs have shown pulsed emission consistent with electron cyclotron maser instability (ECMI) emission from aurorally precipitating electrons \citep{Hallinan2008}. This auroral radio emission, initially interpreted as stellar flaring from brown dwarf LP944-20 by \cite{Berger2001}, has been confirmed many times \citep[e.g.,][]{berger2002, Burgasser2005, Berger2005, Berger2006, Hallinan2006, Hallinan2007, Route2012}. 

Additional evidence for auroral precipitation on brown dwarfs comes from detections of H$\alpha$ emission and potentially from observations of photometric variability. While ongoing radio investigations \citep[e.g.][]{Route2016, Richey-Yowell2020, OrtizCeballos2024, Kao2024} show that of the general population of brown dwarfs only about 7-10\% exhibit observable radio emission, \cite{Pineda2016} defined a sample of brown dwarfs with strong H$\alpha$ emission and \cite{Kao2016} determined that nearly all of them showed radio emission. The presence of H$\alpha$ emission coincident with the radio emission indicates that not only does an energetic electron beam exist to produce ECMI emission, but that it interacts with the brown dwarf atmosphere. In addition, several authors have proposed that observed photometric variability of brown dwarfs could provide additional evidence of auroral processes linked to observed radio emission \citep[e.g.,][]{Hallinan_2015}. Though patchy silicate clouds or dust have been shown to provide one source of photometric variability for warm brown dwarfs \citep{Morley2014, Richey-Yowell2020}, auroral processes may also contribute. This is particularly important for cold brown dwarfs where clouds form below the photosphere and therefore do not contribute to observed variability.

\begin{figure}[ht]
\centering
\includegraphics[width=\columnwidth]{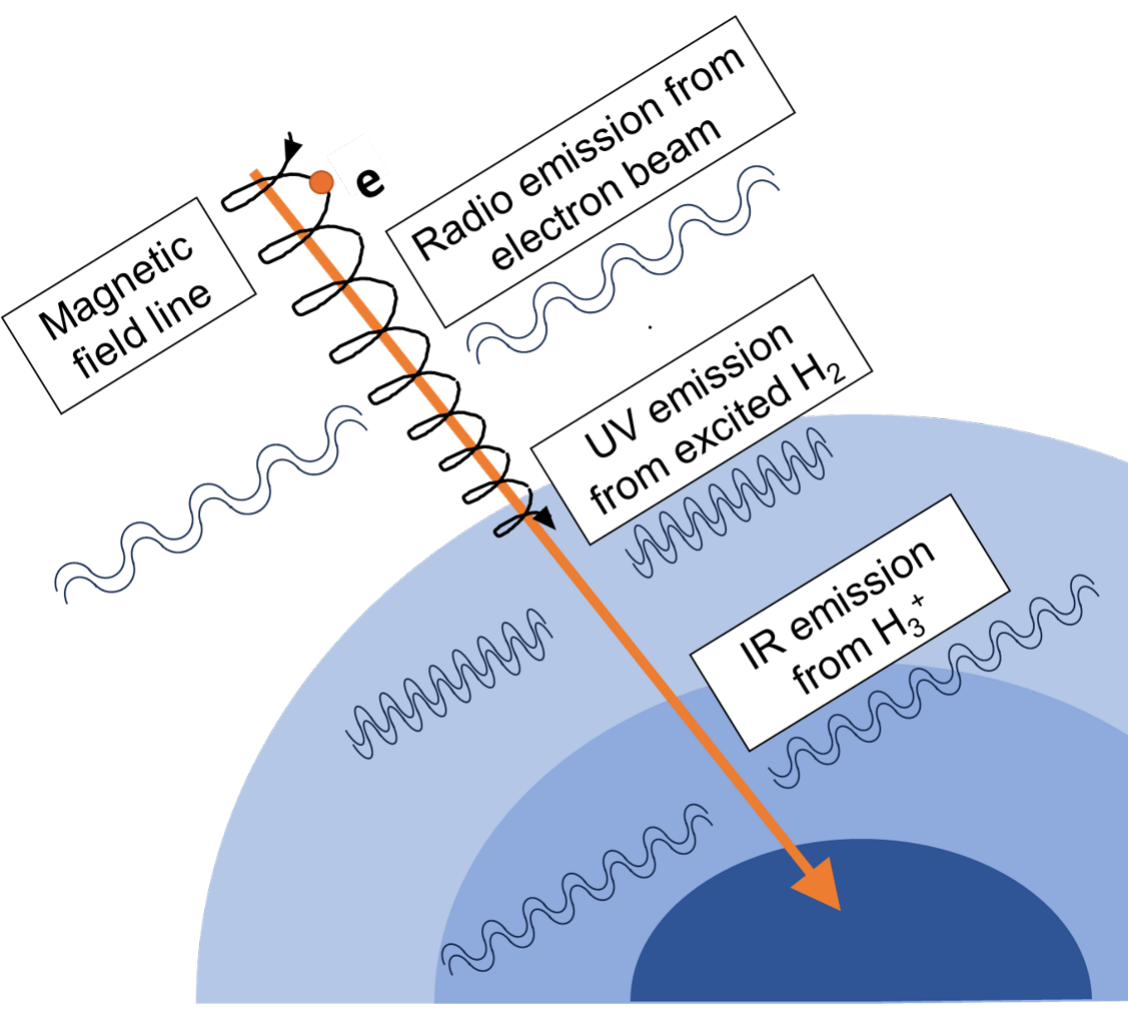}
\caption{A schematic of the emission processes involved in electron-beam aurorae.}
\label{fig:aurora_schematic_jupiter}
\end{figure}

Solar system gas giants provide our closest analogs for understanding what observables to expect from this type of auroral precipitation. Auroral precipitation on gas giants like Jupiter has been shown to be powered by  magnetic processes similar to those on brown dwarfs. For instance, \cite{Hallinan_2015} model the lightcurve of auroral substellar object LSR J1835+3259 to demonstrate that the processes powering magnetic activity are similar to those driven by high energy electron beams incident on the atmosphere of Jupiter. Observations of these solar system analogs suggest that in addition to this radio emission we should expect observable ultraviolet (UV) and infrared (IR), as well as H$\alpha$, emission from the interaction of these auroral electrons with the brown dwarf atmosphere \citep{Badman2015}. Significant observational effort has been made recently to detect the expected corresponding UV and IR auroral emission. Figure \ref{fig:aurora_schematic_jupiter} depicts the expected auroral picture. However, no definitive detection has been reported. For instance, \cite{Saur2018} report UV Hubble Space Telescope (HST) observations of the same object observed by \cite{Hallinan_2015}, and present dramatically (several orders of magnitude) weaker UV emission than expected from auroral emission similar to Jupiter's. In a subsequent paper, \citet{Saur2021} present inconclusive evidence of UV emission from the brown dwarf 2MASS J12373919+6526148. Other surveys searching for auroral IR emission have also resulted in nondetections \citep{GibbsFitzgerald2022, Pineda2024}. As a step towards understanding the missing UV and IR emission, in this work we model the auroral precipitation process.

In addition to the unexplained absence of auroral UV and IR emission, another mystery motivates models of brown dwarf aurorae. Recent intriguing James Webb Space Telescope (JWST) observations indicate the presence of a thermal inversion in the atmospheres of the Y-dwarf CWISEP J193518.59-154620.3 (W1935 for short) \citep{Faherty2024} and the T-dwarf SIMPJ013656.5+093347.3 (SIMP0136 for short) \citep{Nasedkin2025}. Such thermal inversions requires a heating source, which in the solar system can be provided via auroral energy deposition. Recent observations suggest that W1935 is a pair of binary dwarfs \citep{DeFurio2025}. Binary brown dwarfs commonly exhibit auroral radio emission \citep{Kao2022, Kao2025}, motivating auroral modeling of this source. The low temperature ($\sim$350K) of the companion object is insufficient to provide the source of heating for the inversion \citep{DeFurio2025}. The presence of an inversion implies certain observational predictions, and in particular additional spectral features from methane and ammonia \citep{Suarez2025}. Models of auroral precipitation in brown dwarf atmospheres are needed to assess whether auroral heating could provide the missing source of energy implied by these JWST observations and to motivate future observations.

In order to shed light onto these open questions and inform searches for brown dwarf aurorae, we must first develop a better understanding of the mechanism and atmospheric conditions which are likely to power auroral emission on brown dwarfs. As a first step towards modeling the expected emission under different beam mechanisms and atmospheric properties, in this work we develop a model of the interaction of energetic electrons as they propagate through substellar atmospheres.

Several approaches to the problem of simulating electron beam propagation through planetary or substellar atmospheres have been explored in the past. These include the Continuous Slowing Down Approximation (CSDA) method \citep{Gerard1982}, which assumes electrons simply lose energy continuously as they propagate downwards without scattering. A second approach, to directly solve the Boltzmann equation for the phase space density of electrons \citep[e.g.,][]{Lummerzheim1989, Waite1981}, has been applied to gas giants including Saturn \citep[e.g.,][]{Moore2008, Galand2009, Galand2011}. This method has the drawback of becoming very complex when including realistic cross sections and many possible interaction processes.  Another approach is to take a bulk fluid dynamics view. For instance, \cite{Grodent2001} and the more recent works which build from it \citep[e.g.,][] {Gustin2009} use a coupled two-stream electron transport model and 1D thermal conduction model to calculate the upwards and downwards transport of energy through the atmosphere of Jupiter. Other models, such as that employed as part of the Global Airglow (GLOW) model of Earth's aurora, analytically solve two-stream electron transport equations assuming only forward and backward scattering and time-independence to solve for energetic electron flux \citep[e.g.,][]{Solomon2017}. Finally, other authors have explored Monte Carlo type simulations of the interactions of energetic electrons with atmospheres  \citep[e.g.,][]{Cicerone1971}. For instance, \cite{Hiraki2008} consider auroral electron precipitation in the H$_2$ dominated Jovian atmosphere. In these methods, a beam of electrons is considered and followed as they propagate individually, and their interactions with the atmosphere are tracked. This is computationally expensive compared to the analytic approaches described above, but allows for greater flexibility in the physics included in the simulation, as well as the ability to extend the results easily to different atmospheres. 

In this work we apply a Monte Carlo method to the problem of energetic electron beams interacting with brown dwarf atmospheres. We first introduce the model and parameters in Section \ref{sec:model}. In Section \ref{sec:results}, we validate against the results of \cite{Hiraki2008} for Jupiter. We extend their algorithm by allowing for the simulation of different objects and optimizing certain aspects of the algorithm for computational efficiency (for instance, allowing for a time-variable and electron-dependent timestep). We then extend our analysis to a set of substellar objects with properties representing the range of objects observed to produce radio aurorae, $\log_{10}(g)= 4$-$5$ [cgs] and $T_\mathrm{eff}=900$~K-$2000$~K. In Section \ref{sec:Qion} we then present an analytic parameterization of the volumetric interaction rates for various atmospheric processes, which requires only an assumed density profile and electron beam energy spectrum, so that these results can be readily extended to new objects without the need for additional simulations. In Section \ref{sec:discussion} we discuss implications and extensions of our results.

\section{Model} 
\label{sec:model}

We simulated the interactions of a beam of energetic electrons as they propagate through H$_2$ dominated substellar atmospheres. In this section we describe the parameters related to the model and object under study which must be set prior to running a simulation. Code is available on github\footnote{\url{github.com/annazuckerman/auroral_electron_beam_simulation}}.

\subsection{Model Physics and Atmospheric Specification}
\label{subsec:params}

For each simulation we first defined the atmospheric profile. In this work we considered H$_2$ dominated atmospheres, and specified a density profile for each. We allow electrons to interact only with H$_2$, but we note that smaller populations of other constituents such as He, hydrocarbons including CH$_4$, and dust would serve as an unaccounted-for energy drain, thermalizing the electrons faster. \cite{Hiraki2008} find that including the secondary constituent He in their similar simulation for Jupiter changes their results by less than 10\%. As discussed in Section \ref{sec:discussion}, the presence of small densities of hydrocarbons in the atmosphere will be important for future work determining the expected escaping UV and IR emission, but the abundances are negligible for modeling electron precipitation.

We derived density profiles from custom atmospheric models generated with \texttt{PICASO} \citep{Mukherjee2023}, an atmospheric climate model that has been used to generate grids of models \citep{Mukherjee2024}, and analyze JWST observations of substellar objects \cite[e.g.][]{Beiler2023,Miles2023,Lew2024}.
For a given surface gravity and effective temperature, \texttt{PICASO} calculates the 1D thermal structure, mixing ratios, and mean molecular mass as a function of pressure in the atmosphere. The model assumes radiative–convective, hydrostatic, and chemical equilibrium. To compute the converged pressure–temperature profile, the model iteratively solves for the atmospheric structure by updating chemical abundances, opacities, and radiative fluxes, allowing for multiple radiative and convective zones, until radiative–convective equilibrium is achieved. Vertical mixing in the atmosphere is described by $K_{\rm zz}$, the eddy diffusion coefficient. In the convective regions of the atmosphere, this is computed using mixing-length theory \citep{GieraschConrath1985}, while in the radiative zones, we follow the prescription described in \citet{Moses2022}. \texttt{PICASO} does not consider molecular diffusion. We assume that turbulent mixing, constrained by $K_{\rm zz}$, dominates all the way up to the top of the atmosphere. 

The \texttt{PICASO} atmospheric profile extends to the homopause, or a pressure of about $10^{-6}$ bars. In this region, we derived altitude for a given pressure level as the integral of $1/\rho g$, where $\rho$ is mass density, assuming an ideal gas and defining the zero-point in altitude at a pressure of one bar, and letting gravity change with altitude. We did this by interpolating the outputs of the \texttt{PICASO} atmospheric models in logarithmic space onto a fine grid of 100,000 points, and numerically integrating. Then, the density of H$_2$ was defined using the ideal gas law and H$_2$ mixing ratio. 

\begin{figure}[ht]
\centering
\includegraphics[width=\columnwidth]{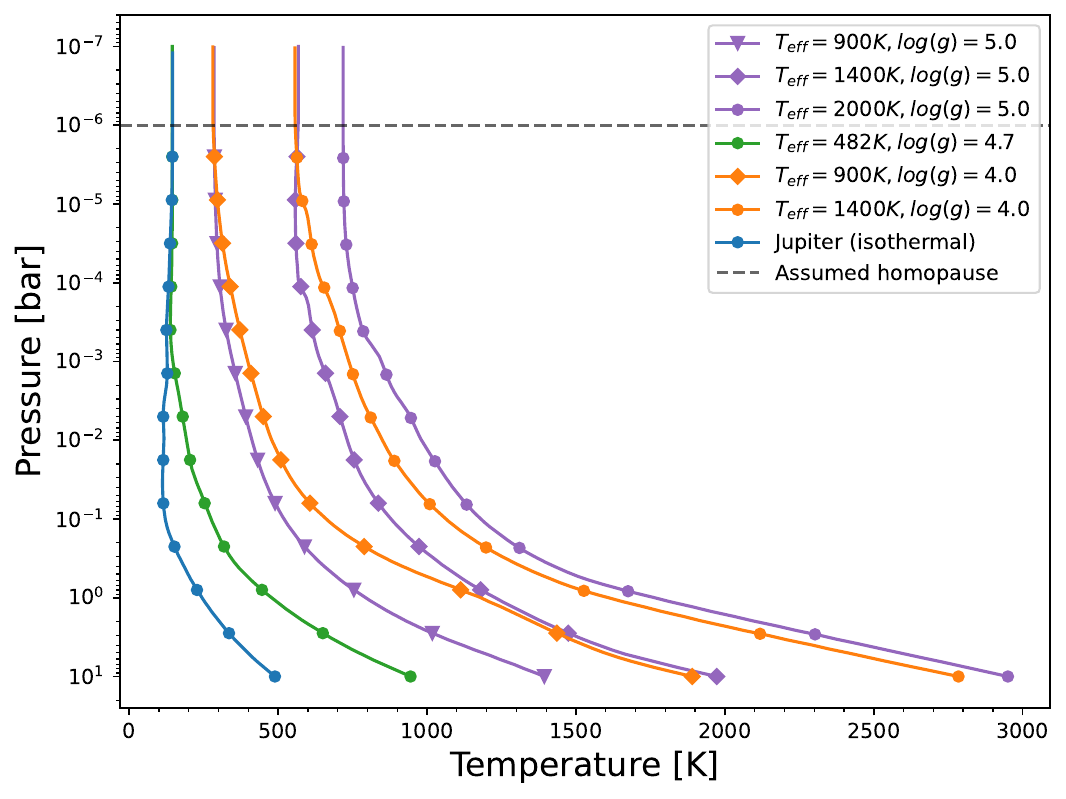}
\caption{Temperature profiles for the objects considered in this work. For Jupiter, the isothermally extended profile is shown.}
\label{fig:PT}
\end{figure}

\begin{figure*}[ht]
\centering
\includegraphics[width=\textwidth]{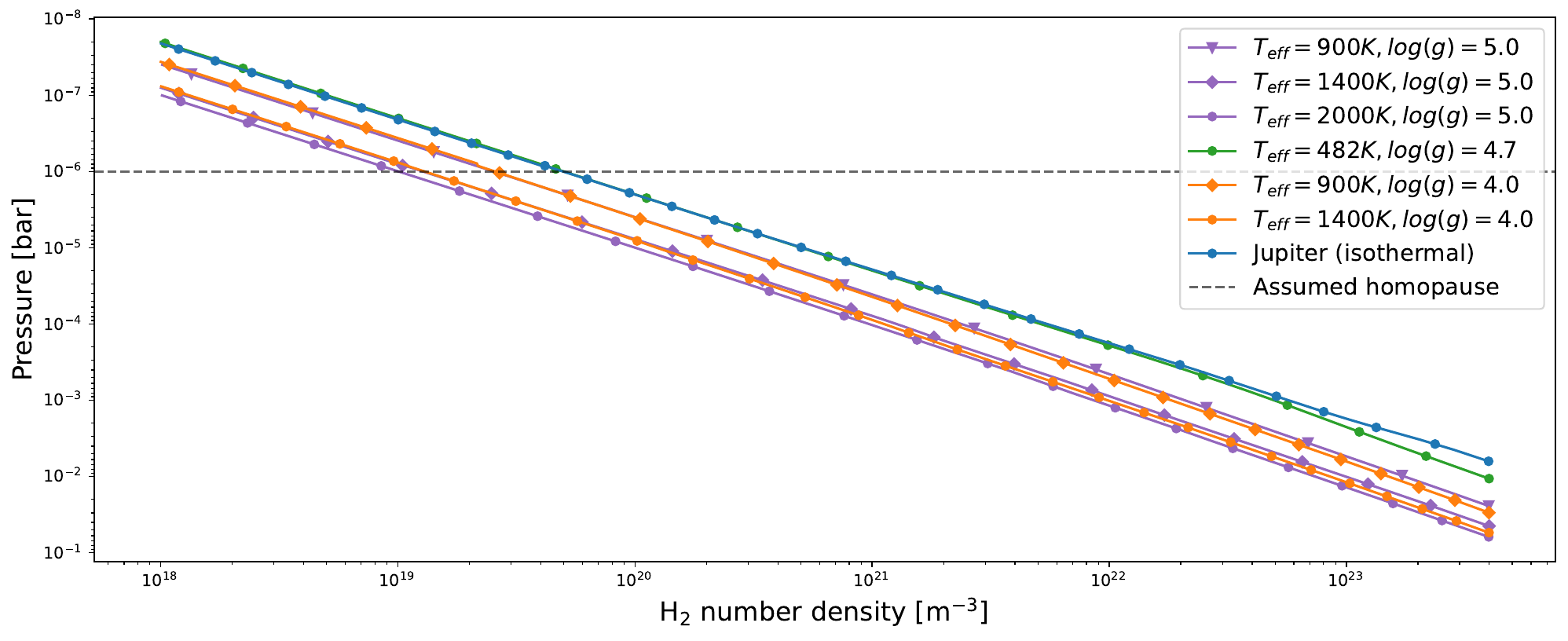}
\caption{Density profiles for the objects considered in this work. For Jupiter, the isothermally extended profile is shown.}
\label{fig:density_profiles}
\end{figure*}

Above the homopause, each atmospheric constituent has its own scale height and the physics assumed in \texttt{PICASO} no longer holds. Many of the interactions we are interested in occur at lower pressures, and so above the homopause we extended the atmosphere isothermally. Though real atmospheres are likely not perfectly isothermal above the homopause, the \texttt{PICASO}-generated temperature profiles level off at high altitudes as they approach the homopause, motivating this assumption (as shown in Figure \ref{fig:PT}). Deviations from isothermal above the homopause would impact the presented profiles in altitude space, but importantly would not significantly impact the derived parameterization, as discussed in Section \ref{subsec:parameterization}. We took the density profile in this isothermal region to decay from the value at the homopause as governed by hydrostatic equilibrium. For Jupiter, instead of extending the atmosphere isothermally above the homopause we used data from the Galileo probe at high altitudes where it exists \citep{Sief1997}. Density can vary greatly between objects at the same pressure level due to the wide range of effective temperature and surface gravity we considered, as shown in Figure \ref{fig:density_profiles}. 

Next, we defined what interactions an energetic electron is permitted to undergo in our atmosphere. Following the basic procedure outlined in \cite{Hiraki2008}, we considered the ten dominant interactions: 1) rotationally elastic scattering, 2) ionization, 3) B electronic excitation, 4) C electronic excitation, 5) a electronic excitation, 6) b electronic excitation, 7) c electronic excitation, 8) e electronic excitation\footnote{Letters are labels for different excited states of H$_2$. Uppercase letters denote excitations to singlet states while lowercase denotes excitations to triplet states. For instance, B and C denote the Werner and Lyman band system upper levels respectively. See \cite{Scarlett2021} for a description of these states.}, 9) vibrational excitation, and 10) rotational excitation. Other interactions, such as other types of excitations, dissociation, and recombination, have small cross sections, and would serve as a small further energy drain. We took cross sections from the MCCC online database \citep{Scarlett2020, Scarlett2023} for all processes but ionization, which we took from the National Institute of Standards and Technology (NIST) database \citep{Kim2004}. We show cross sections in the Appendix. At the temperatures we considered (tens to thousands of Kelvin), the Boltzmann equation for the partition of states requires that H$_2$ is almost entirely in it's ground vibrational state. However, a non-negligible fraction of atmospheric H$_2$ starts in a higher rotational state. For this reason, for rotational excitation we determined a cross section for excitation from each of the initial states that have greater than $1\%$ of the H$_2$ population at the effective temperature of the atmosphere. A 1\% threshold was chosen to balance capturing the bulk of states with practical feasibility; inclusion of the next few most populated states does not meaningfully change the cross section. We then took the overall rotational excitation cross section to be a linear combination of these cross sections, weighted by the population fraction of H$_2$ starting in each initial state. A minor effect of our assumption that the atmosphere is isothermal above the homopause could be through these assumed rotational cross sections, for which the distribution of initial states does depend on temperature. We took the cross sections for each of these ten interactions under a non-relativistic assumption. While this approximation holds for low energy electrons, it breaks down for higher energies. The rest energy of an electron is approximately 511 keV, comparable to the highest kinetic energy of electrons in our simulation. Thus, for a more accurate treatment the cross sections and kinematics should be adjusted for the relativistic case. Results presented here for the highest energies should be considered approximate, but are included for comparison to literature results which treat high energies non-relativistically \citep{Hiraki2008}.

We then defined the energy lost by the incident electron (and gained by any secondaries) for each type of interaction. We assumed motion of the H$_2$ is negligible during the interaction due to its dramatically larger mass. We took the ionization energy partition from \cite{Vahedi1995}, and each excitation energy from the MCCC database. Secondary electron energy is strongly peaked and has a maximum value set by the energy remaining after the energy lost to ionization. For elastic scattering all energy is by definition retained by the incident electron. We took the scattering angles for ionization and elastic scattering to be given by the screened Rutherford formula \citep{Lummerzheim1989}, and scattering to be isotropic for the other interactions, after \cite{Hiraki2008}. The Rutherford scattering angle distribution is more strongly peaked towards forward scattering angles for increasing incident electron energies.

\subsection{Model Algorithm}
\label{subsec:algorithm}
We followed the general scheme described by \cite{Hiraki2008}, with modifications for efficiency and for our scientific cases of interest. 

First, for each run we defined the incident electron beam energy, orientation, and intensity, as well as the atmospheric properties and interaction cross sections at the top of the atmosphere. We considered a monoenergetic beam with 1000 incident electrons, but because the simulation is fully parallel with respect to individual electrons, any arbitrary beam spectrum could be approximated through multiple simulation runs varying the initial beam energy. In our parameterization of the total volumetric interaction rate (Section \ref{subsec:parameterization}), we allowed for any arbitrary beam energy spectrum. The number of incident electrons is chosen to balance computational efficiency with ensuring that even the processes with the smallest cross sections exhibit interactions under the most unfavorable conditions.

We defined an initial timestep such that the total interaction probability is less than 0.1 for each electron, with the total interaction probability during a step given by

\begin{equation}
\label{eq:P(int)}
P = 1 - e^{-\sigma_{tot}(\varepsilon) N} \; , 
\end{equation}

\noindent where $\sigma_{tot}$ is the total interaction cross section (see Appendix \ref{sec:xsecs}), $\varepsilon$ is the electron energy, and $N$ is the column number density along the step. 

Then, the main loop of the algorithm is as follows for each timestep: 

\begin{enumerate}[label=(\arabic*)]
    \item \textbf{Define interaction probability:} For each iteration, the algorithm defines the probability of an interaction for each electron. We do this by first using the velocity, angle, and timestep value calculated previously to define a preliminary spatial step. We then determine the column density across this step. The density profile is not an analytically integrable function, and to numerically integrate would be computationally expensive. Instead we approximate the column density over the step by taking the change in density to be linear over the small step size, and we verify that the error in this approximation is less than 0.5\% even for the large steps taken at the top of the simulation. Using the column density, we are then able to define an overall interaction probability for the step, and to generate a random number to choose whether or not an interaction occurs for each electron based on this probability. 
    \item \textbf{Determine interaction location:} For electrons which had an interaction in a given step, we pick a height along the step at which the interaction occurred. The interaction location is chosen using the same random number (in [0,1]) generated previously such that the probability increases with density along the step. This is done by setting this random number equal to the cumulative interaction probability along the step and solving for the position of the interaction.
    \item \textbf{Determine interaction type:} For electrons which interacted, we then pick which interaction occurred, using the relative cross sections for the different interaction types. This is done by generating a new random number and comparing it against the cross sections.
    \item \textbf{Create secondaries:} We then create any secondary electrons produced via ionization and add them to the simulation.
    \item \textbf{Update energies and trajectories:}  Next, we update the energies and trajectories of the electrons. We record only the vertical component and the horizontal component of the electron position and velocity, rather than both components in the horizontal plane, resulting in a 1.5D model. When an interaction occurs, we take the actual height of the interaction, rather than the termination point of the preliminary step, as the starting point of the trajectory going into the next step. We then update the cross sections at that height for use in determining the interaction probability in the next iteration.
    \item \textbf{Remove low energy electrons:} Finally, we remove from the simulation any electrons which reach an energy threshold of 4.48 eV such that they are below the threshold for all but rotational and vibrational excitations. These low energy electrons are subsequently treated in a secondary loop for computational efficiency.
    \item \textbf{Update timestep:} Using the same methodology as described for the initial timestep, we set the new timestep for each electron's current energy and altitude. Our implementation does not use a constant timestep. Instead, at each step and for each electron individually, we set a timestep such that the interaction probability remains below 0.1 always. This allows us to set different timesteps for electrons which may have very different energies, and thus to increase both the accuracy of the simulation (by allowing for smaller timesteps when the probability of an interaction is high), and the efficiency of the simulation (by allowing for longer timesteps when the probability of an interaction is low). 
\end{enumerate}

For each simulation we iterate through these steps until all electrons are low enough in energy such that only rotational and vibrational excitations may occur. The very low velocities of these electrons allow us to dramatically increase computational efficiency by neglecting their small motions and changes in trajectory. These low energy electrons are then allowed to undergo rotational and vibrational excitations until reaching the vibrational excitation threshold (about 0.89 eV), at which point they undergo only rotational excitations until they reach the rotational excitation threshold (about 0.44 eV) and are considered thermalized. The remainder of their energy goes to heating.

\section{Results}
\label{sec:results}
We report eight simulations using 0.1, 0.5, 1, 5, 10, 50, 100, and 500 keV incident beams, validating against previously published results from Jupiter before applying the simulation to a range of other substellar objects. 

The primary quantity of interest which can be derived from our simulation results is the volumetric interaction rate for each type of interaction. We are particularly interested in the ionization interaction rate $Q_{ion}$ (number ionizations m$^{-3}$ s$^{-1}$) because ionizations should produce the expected IR emission, and we discuss in Section \ref{subsec:parameterization} how our analysis extends to other interactions. The quantity directly derived from our simulations is the ionization profile $q_{ion}$ (number ionizations m$^{-1}$ per incident electron). $Q_{ion}$ can be written in terms of the ionization profile $q_{ion}$ 

\begin{equation}
\label{eq:Qion}
    Q_{ion}(z) = \int q_{ion}(\varepsilon_0, z) F(\varepsilon_0) d\varepsilon_0 \; ,
\end{equation}

\noindent where $\varepsilon_0$ is the incident electron energy, $z$ represents altitude within the atmosphere (here defined such that $P(z=0) = 1$ bar) and $F(\varepsilon_0)$ is the energy spectrum of the incident beam (number of electrons m$^{-2}$ s$^{-1}$ eV$^{-1}$). In each simulation we selected $F(\varepsilon_0)$ as a delta function at the energies considered and so picked out specific values of the integral. As discussed in Section \ref{subsec:parameterization}, this allowed us to fit analytic expressions for $q_{ion}$ for each object and beam energy, and thus to determine $Q_{ion}$ for any arbitrary incident beam energy spectrum. We also present interaction rates analogously for the different types of interactions considered, as well as for thermal heating.

\subsection{Jupiter Validation Case}
\label{subsec:Hiraki Jupiter}
First, as a validation for our model, we reproduced the results of \cite{Hiraki2008} for the ionization profile for Jupiter.

\begin{figure}[ht]
\centering
\includegraphics[width=\columnwidth]{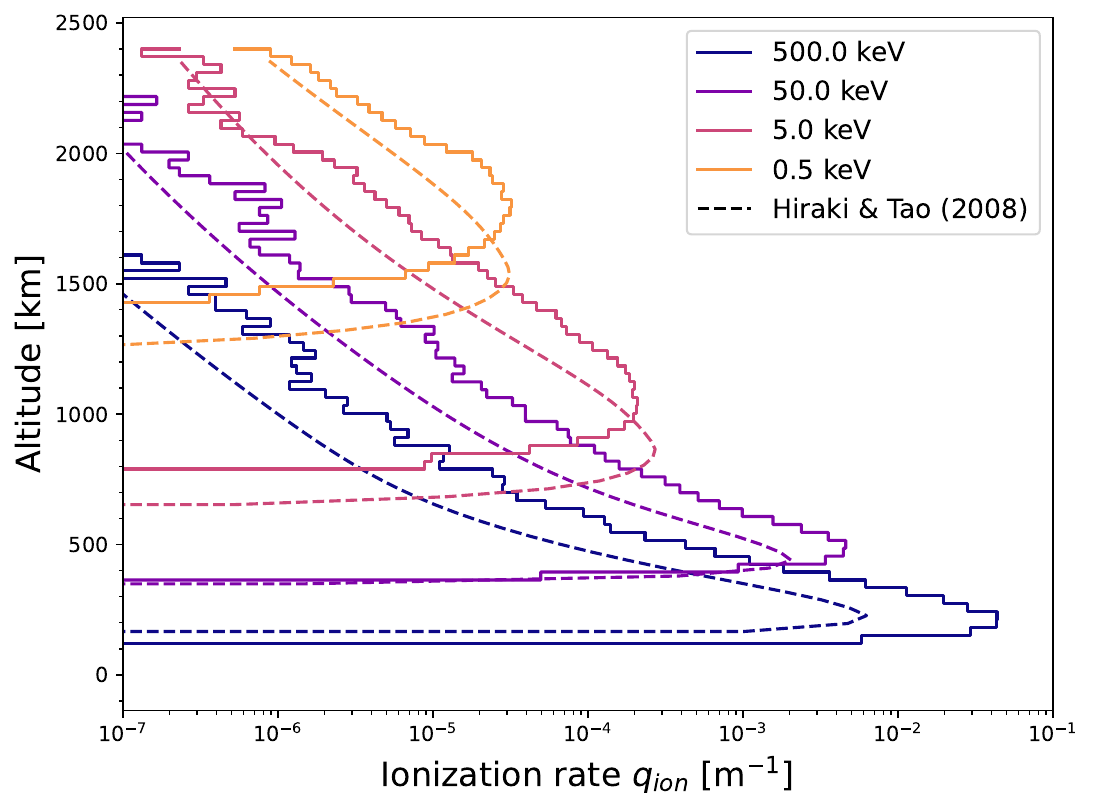}
\caption{Our simulated ionization profile (solid) for Jupiter using the density profile used by \cite{Hiraki2008}, compared with the parameterized ionization profile presented in that paper (dashed). Note that in this and other figures, partial pressure of H$_2$ is reported rather than full atmospheric pressure. This is because in extending the density profile of H$_2$ isothermally, we do not calculate density profiles or pressures for other species and thus for consistency with the isothermal region report partial pressure across the atmosphere. However, this is a small effect on the value of the pressure axis reported, as the mixing ratio of H$_2$ is several orders of magnitude larger than any other species throughout the atmospheres which we consider.}
\label{fig:Jupiter_with_Hiraki_profile}
\end{figure}

Using the density profile given by \cite{Hiraki2008}, we were able to reproduce the qualitative behavior of the published ionization rate curve well (see Figure \ref{fig:Jupiter_with_Hiraki_profile}). The remaining discrepancy between our results and the curve published by \cite{Hiraki2008} is likely explained by our use of updated interaction cross sections compared to those available to these authors, but we do not have access to their cross sections to confirm this. We used an ionization cross section that is overall larger, and in particular larger at the higher initial beam energies, compared with the cross section used by \cite{Hiraki2008} (Figure 12a in their paper), allowing more ionizations to occur higher in the atmosphere and thus shifting the profile to higher altitudes. Our simulations nonetheless reproduced the qualitative behavior seen in the literature, and by updating the physics we were able to extend their previous results. Importantly, \cite{Hiraki2008} report a parameterization for the ionization rate only, allowing us to validate for ionization only. As discussed in Section~\ref{sec:discussion}, producing a parameterization for the other interactions is a key part of this work because of their importance for potentially observable auroral emission.

After validating our simulation against this previous result, we implemented a density profile derived from custom \texttt{PICASO} model outputs and Galileo probe data. This density profile differs from that used by \cite{Hiraki2008}, and results in an atmosphere which is less dense at higher altitudes and more dense at lower altitudes compared with the model adopted by \cite{Hiraki2008}. We chose to use our derived density profile both in order to take advantage of modern atmospheric modeling results, and for consistency with our other science case. Using this density profile, we ran the simulation with a range of incident electron energies from 0.1 keV to 500 keV. At the high-energy end of this range, the results should be considered approximate because the non-relativistic assumption we used for incident particles begins to break down. 

\begin{figure}[ht]
\centering
\includegraphics[width=\columnwidth]{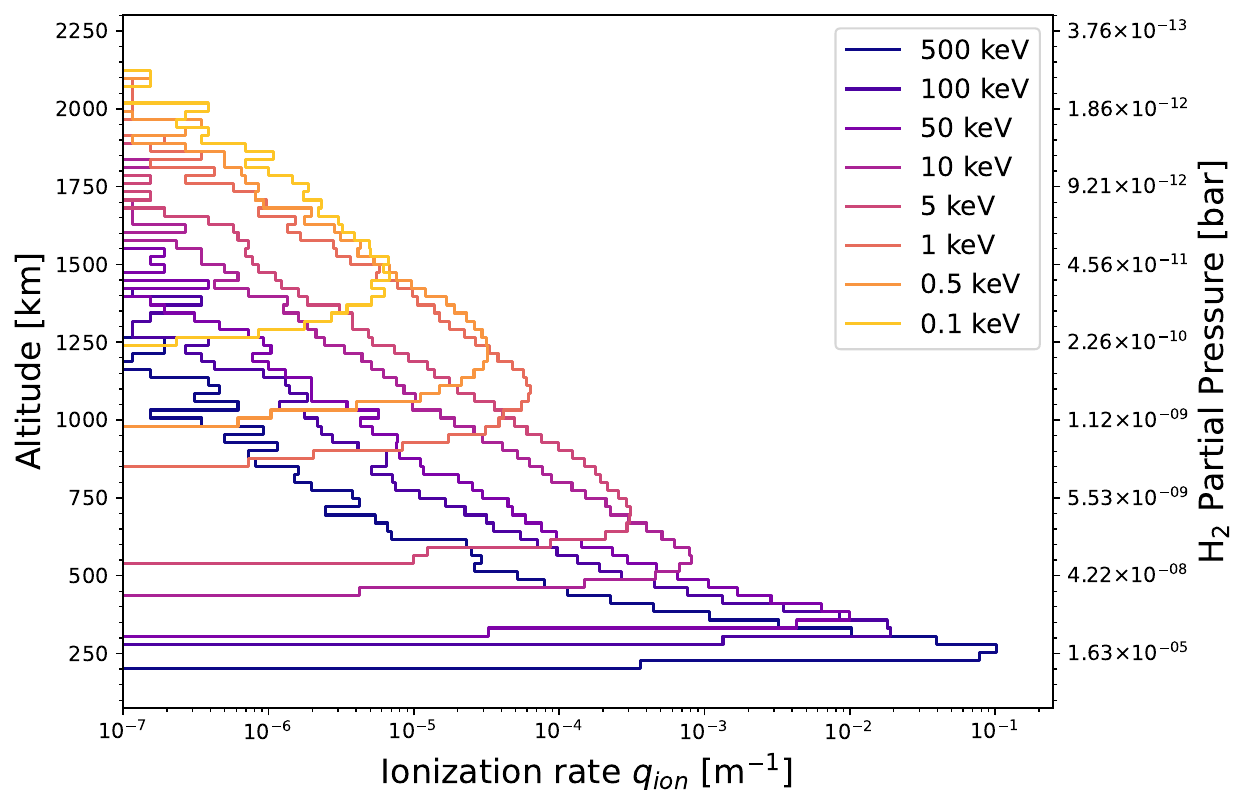}
\caption{Ionization rate per incident electron as a function of altitude or pressure for Jupiter using the density profile derived from custom \texttt{PICASO} model outputs and Galileo measurements, for a range of incident electron energies. As described in Section \ref{subsec:params}, the relationship between altitude and pressure is determined from the outputs of a custom \texttt{PICASO} model plus an isothermal extension.}
\label{fig:Jupiter_compare_energies}
\end{figure}

As we show in Figure \ref{fig:Jupiter_compare_energies}, ionization rate peaks deeper in the atmosphere for higher energy incident electrons. This shift to lower altitudes is due to the lower interaction cross section for more energetic electrons, requiring higher densities before interactions become significant, and the fact that higher energy electrons require more interactions to thermalize. The vertical width of the peaks is due to the probabilistic nature of the electron precipitation, with some electrons undergoing their first ionization higher up and some lower down. The ionization rate also becomes more strongly peaked with increasing electron energy, and is shifted to higher ionization rates because higher energy electrons have more energy to lose to interactions with the atmosphere.

\begin{figure}[ht]
\centering
\includegraphics[width=\columnwidth]{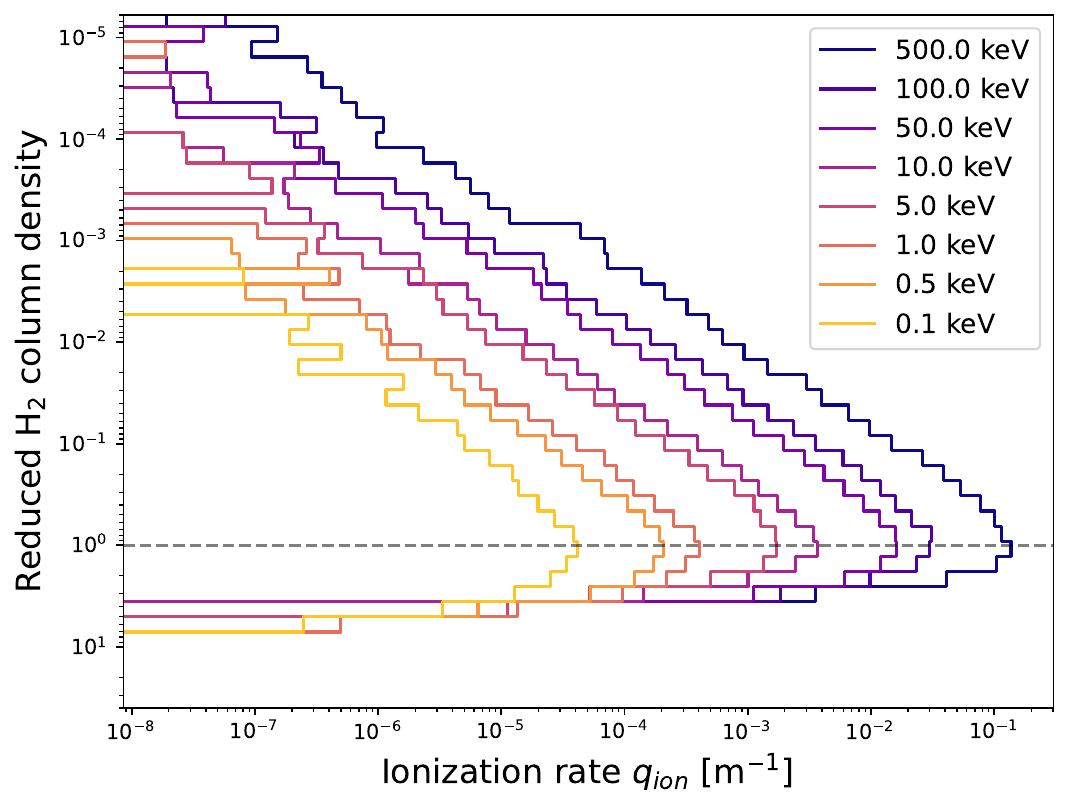}
\caption{Ionization rate per incident electron as a function of reduced column density $N/N_0$ for Jupiter using the density profile derived from custom \texttt{PICASO} model outputs and Galileo data, for a range of incident electron energies. Ionization profiles for different energies align in this space. The dashed line is at a reduced column density of 1.}
\label{fig:Jupiter_compare_energies_R}
\end{figure}

Because column density, rather than altitude, is the underlying quantity which governs where the ionization profile peaks and falls off, we scale our profiles by the column density of the peak ionization rate. In particular, we can plot the ionization profile as a function of the reduced column density $N/N_0$ where $N$ is the column number density and $N_0$ is the column density at the peak of the profile. We used kernel density estimation to provide a more precise estimate of the peak location than our histogram binwidth, as described in Section \ref{subsec:evaluation}. Because the ionization cross section is lower for higher energies, more energetic electrons must reach higher atmospheric densities for significant interactions to occur, and thus must traverse through a greater column density before the ionization rate peaks compared to lower energy electrons. Normalizing by $N_0$ removes this energy-dependent effect. In Figure \ref{fig:Jupiter_compare_energies_R}, we compare ionization rates as a function of $N/N_0$. In this space, the peaks align in location and the profiles across energies are self-similar. In Section \ref{subsec:new_cases}, we demonstrate that an analogous energy dependence of the peak location exists for other objects, and that expressing the peak location in column density eliminates the difference that arises due to the different density profiles of the objects.  

\begin{figure}[ht]
\centering
\includegraphics[width=\columnwidth]{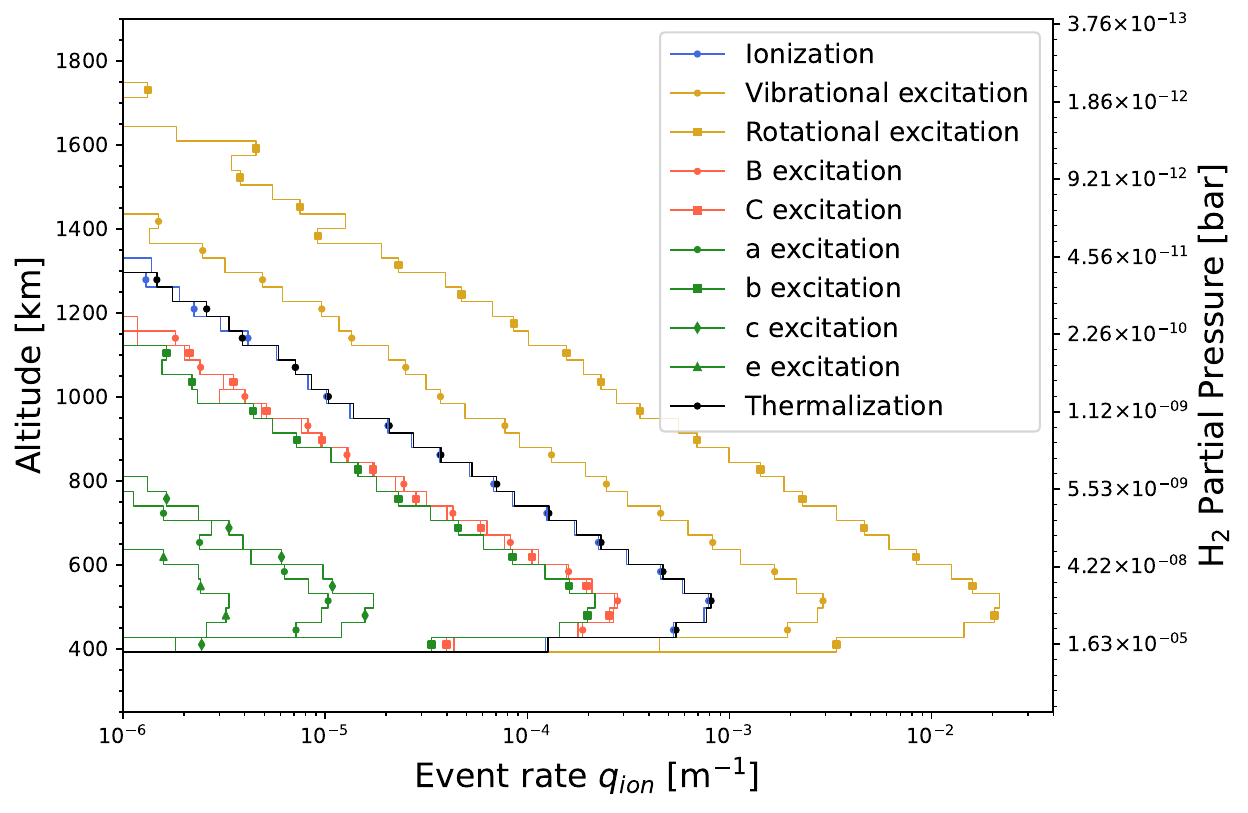}
\caption{Rates for ionization compared with other interaction types, for a 10 keV incident electron beam for Jupiter. The profiles are similarly shaped due to the fact that other cross sections are small until an ionization occurs to produce lower energy electrons, which then scatter in the neighborhood of the original ionization. Profiles are scaled by the relative strengths of the cross sections, and interactions with similar cross sections have profiles that fall close together. For instance, the B and C excitation profiles are very similar due to their similar cross sections.}
\label{fig:compare_events}
\end{figure}

\begin{figure*}[ht]
\centering
\includegraphics[width=\textwidth]{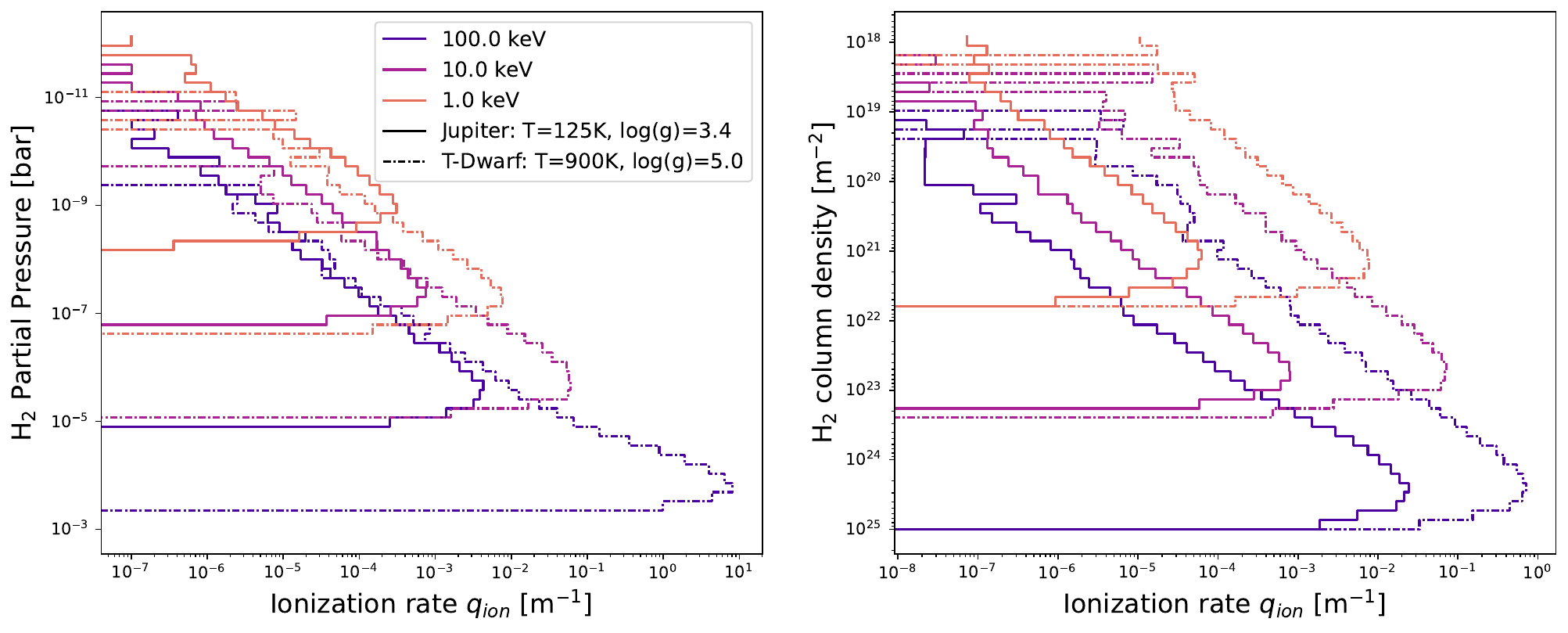}
\caption{Left panel: Ionization rate per incident electron as a function of pressure for Jupiter and the T-dwarf for several incident electron beam energies. Ionization rate peaks at significantly higher pressures and with an overall greater number of ionizations for the T-dwarf due to the effect of higher gravity to condense the atmospheric profile. Right panel: Ionization rates for the two objects align when examined in column density space.}
\label{fig:compare_Jup_900K_P}
\end{figure*}

\begin{figure*}[ht]
\centering
\includegraphics[width=\textwidth]{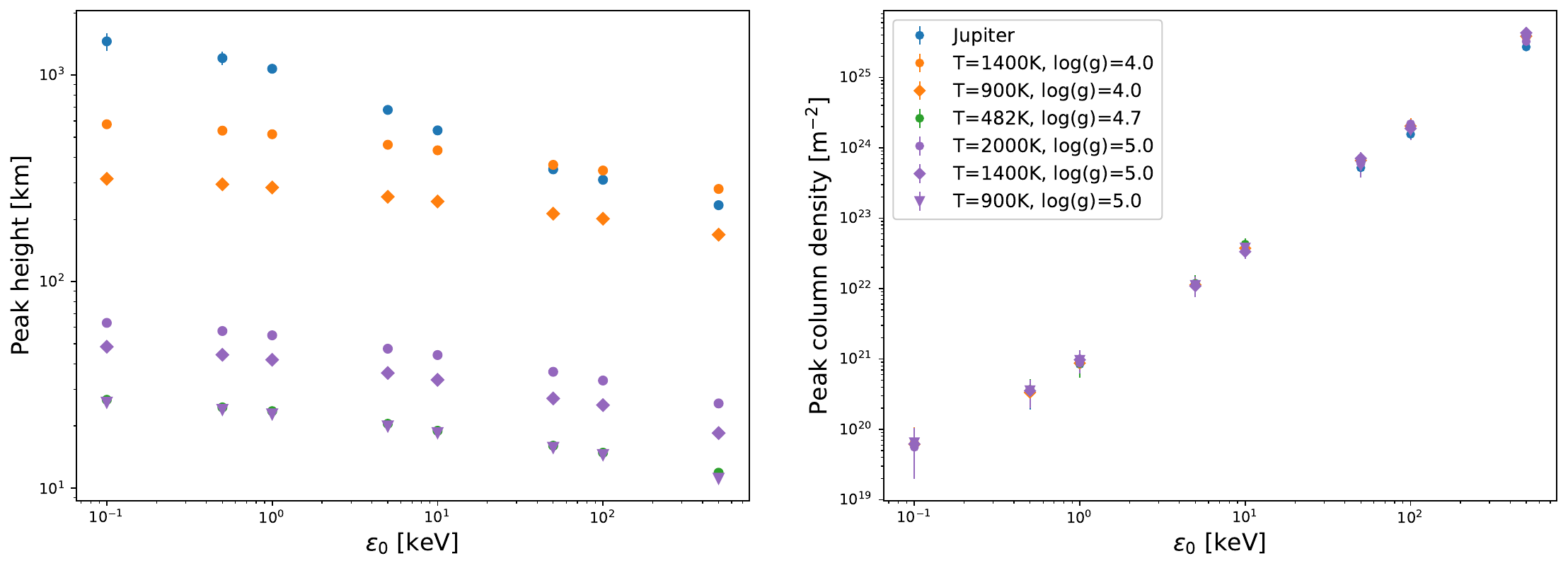}
\caption{Left panel: Ionization peak altitude as a function of incident beam energy. Note that the purple triangles almost entirely overplot the green circles, due to the competing effects of gravity and temperature for these two objects. Right panel: Ionization peak column density as a function of incident beam energy. 1$\sigma$ errorbars are shown, and are calculated as described in Section \ref{subsec:evaluation}. Dramatic differences in peak location in altitude collapse together when examined in column density space.}
\label{fig:peak_z_and_N}
\end{figure*}

\begin{figure*}[ht]
\centering
\includegraphics[width=\textwidth]{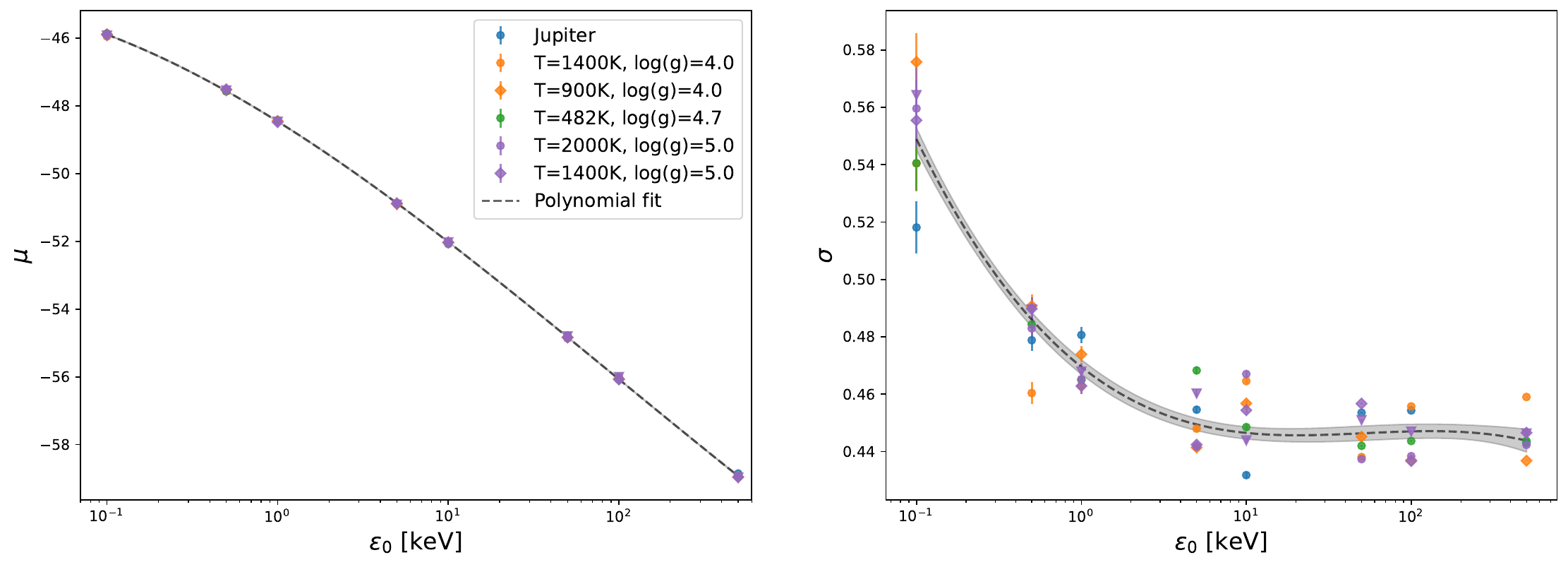}
\caption{Polynomial fits to the parameters $\mu$ and $\sigma$ of the Moyal distributions which fit the distribution of $-\ln(N)$ of ionizations. One-standard deviation errorbars on datapoints and uncertainty regions on the dashed fits are shown (the uncertainty region on the left-hand plot is not visible).}
\label{fig:mu_and_sigma}
\end{figure*}

We also recorded the altitudes of the nine other interactions, plus thermalization of electrons. We find that all of these rates track closely with ionization rate as a function of altitude, though scaled in intensity due to the very different cross sections, as shown in Figure \ref{fig:compare_events}. This is because at high energies ionization dominates, and once an ionization has occurred the resulting cascade of secondaries are unlikely to move far vertically from the location of the ionization due to being low energy and thus slow moving. Secondaries typically are produced with low energies (even for the highest energy primaries we consider, secondaries are produced with energies higher than 10 eV for less than 50\% of ionizations), meaning that secondary electrons are slow moving and secondaries are produced near or below the thresholds for most interactions. 

\subsection{Comparison Across the Brown Dwarf Regime}
\label{subsec:new_cases}
After validating our model on the case of Jupiter, we extended our results to a broader set of brown dwarf cases. Producing a model which is generalized across all H$_2$ dominated atmospheres is a key result of this work.
We defined a new atmospheric density profile for each case based on custom \texttt{PICASO} model outputs as described above, plus an isothermal extension above the homopause. For comparison, we also simulated a Jupiter-like case using this isothermally extended atmospheric model rather than the measured Galileo data at high altitudes. We simulated six objects to fill out the parameter space and to simulate specific observed objects as displayed in Table \ref{tab:object_params}.

\begin{table}[]
\caption{The range of simulated objects presented in this work.}
\begin{tabular}{ccc}
\textbf{Object \textbf{spectral type}} & \textbf{log(g) {[}cgs{]}} & \textbf{T$_{eff}$ {[}K{]}} \\ \hline
T-dwarf         & 5.0                       & 900                        \\
L-dwarf         & 5.0                       & 1400                       \\
L-dwarf         & 5.0                       & 2000                       \\
 Y-dwarf (W1935) & 4.7                       & 482                        \\
T-dwarf         & 4.0                       & 900                        \\
L-dwarf         & 4.0                       & 1400                       \\
Jupiter         & 3.4                       & 125                       
\end{tabular}
\label{tab:object_params}
\end{table}

As in the Jupiter case, ionizations occur deeper in the atmosphere for higher incident beam energies for all objects (see the right panel of Figure \ref{fig:compare_Jup_900K_P} for a comparison of Jupiter and a representative T-dwarf). However, the locations of the peaks in altitude and pressure level differ substantially across objects, because of the very different density profiles. The differences in ionization rate profiles across objects are driven directly by the differences in density profile due to changing gravity and temperature. The primary driver is gravity, with higher gravity objects having density profiles which are more condensed spatially (shorter scale heights), resulting in atmospheric interactions occurring lower down in altitude. In addition, higher temperature objects have puffier atmospheres, or more extended density profiles, resulting in atmospheric interactions occurring higher up in altitude. In the left panel of Figure \ref{fig:compare_Jup_900K_P}, we compare the ionization rates for Jupiter and a representative T-dwarf in pressure space. The larger peak in ionization rate for the T-dwarf is due to the greater localization of interactions under the more condensed density profile for that atmosphere, rather than a greater overall number of ionizations. The fact that the magnitude of $q_{ion}$ appears larger for the T-dwarf is due to the effect of binning by altitude and plotting against pressure or column density for different density profiles. There is less altitude represented in each plotted bin for the more condensed density profile of the higher gravity T-dwarf. Looking at the profiles in column density space brings the two cases closer together (see the right panel of Figure \ref{fig:compare_Jup_900K_P} and the right panel of Figure \ref{fig:peak_z_and_N}).

Despite very different density profiles resulting in very different interaction rates in altitude, we can see that when examined in column density space rather than in altitude the interaction rates collapse together across objects. We show this for the full range of objects studied in Figure \ref{fig:peak_z_and_N}, which displays the locations of the peaks as a function of energy. This fact is the basis of the parameterization discussed in Section \ref{subsec:parameterization}. 

\subsection{Analytic Parameterization}
\label{subsec:parameterization}
We have reported here profiles in the form of histograms for ionization, excitations, and thermal heating. We have demonstrated visually and qualitatively how these profiles scale with altitude and with incident electron beam energy. Next, we determined an analytic parameterization for these profiles, so that our results may be more easily extended to a wider range of conditions and in order to carry out the integral in Equation \ref{eq:Qion} for any arbitrary electron beam energy spectrum. Our task was to determine an analytic expression for $q_{ion}$ as a function of altitude (through column density) for any incident beam energy which holds across objects. Note that in deriving the parameterization we simulated Jupiter with the density profile isothermally extended above the domain of the \texttt{PICASO} model rather than extended with Galileo data for consistency across objects, and in Section \ref{subsec:evaluation} we verify that the parameterization nonetheless performs well on Jupiter simulated using Galileo data.

In order to derive a functional form for $q_{ion}$, we first determined a probability distribution function $\mathcal{P}$ for our ionization rate. Though we could have achieved a similar parameterization by fitting a functional form such as a series of polynomials to $q_{ion}$, finding an approximation of the distribution provided better physical insight into the shape of the distribution. We explored many possible distribution functions matching the observed qualitative behavior, including those in the Gamma and Generalized Extreme Value families. We found that a Moyal distribution, which is a transformation of a Gamma distribution, matches well the distribution of ionizations in $-\ln(N_{ion})$ space, where $N_{ion}$ is the column density at which the ionization occurs. The Moyal distribution is described by location parameter $\mu$ and scale parameter $\sigma$, following the general form $\mathcal{P}(x|\mu,\sigma) = \frac{1}{\sqrt{2\pi}\sigma}\exp[{-\frac{1}{2}(z + \exp(-z)]}$, where $z = (x - \mu)/\sigma$. The use of a Moyal distribution to model ionization rate is motivated by its original derivation as a description of energy loss due to ionizations from charged particle collisions \citep{Moyal1955}. We can then transform the Moyal distribution for the probability distribution of $-\ln(N_{ion})$, to a distribution for $N_{ion}$:

\begin{equation}
\label{eq:pdf}
    \mathcal{P}(N_{ion}) = \frac{1}{\sigma\sqrt{2\pi}}N_{ion}^{\frac{1}{2\sigma} - 1}\exp\left[\frac{\mu }{2\sigma } - \frac{1}{2}N_{ion}^{1/\sigma} e^{\frac{\mu }{\sigma }}\right]
\end{equation}

We used the Bayesian framework of the \texttt{PyMC} \citep{pymc} package to estimate the most likely parameters $\mu$ and $\sigma$ of the Moyal distribution fit to each simulation run across eight energies and seven objects. We examined these parameters as functions of energy and of effective temperature and surface gravity, and found that across objects they were well fit by continuous polynomials of the incident electron beam energy alone. We found good fits for $\mu$ and $\sigma$ to be 
\begin{equation*}
\mu = -48.459 -1.366\ln(\varepsilon_{0}) -  0.094\ln^2(\varepsilon_{0}) + 0.007\ln^3(\varepsilon_{0})
\end{equation*}

\begin{equation}
\sigma = 0.470 - 0.020\ln(\varepsilon_{0}) + 0.005\ln^2(\varepsilon_{0})  - 0.0004\ln^3(\varepsilon_{0}) 
\end{equation}

\noindent where $\varepsilon_0$ is the incident beam energy in keV, and both $\mu$ and $\sigma$ have the same units as the fit quantity, $-\ln(N_{ion})$ where $N_{ion}$ is in m$^{-2}$.
In Figure \ref{fig:mu_and_sigma}, we show that $\mu$ and $\sigma$ are primarily a function of energy across objects. The residuals of these fits for $\mu$ and $\sigma$ do not show trends in the object properties surface gravity and effective temperature.

\begin{figure}[ht]
\centering
\includegraphics[width=\columnwidth]{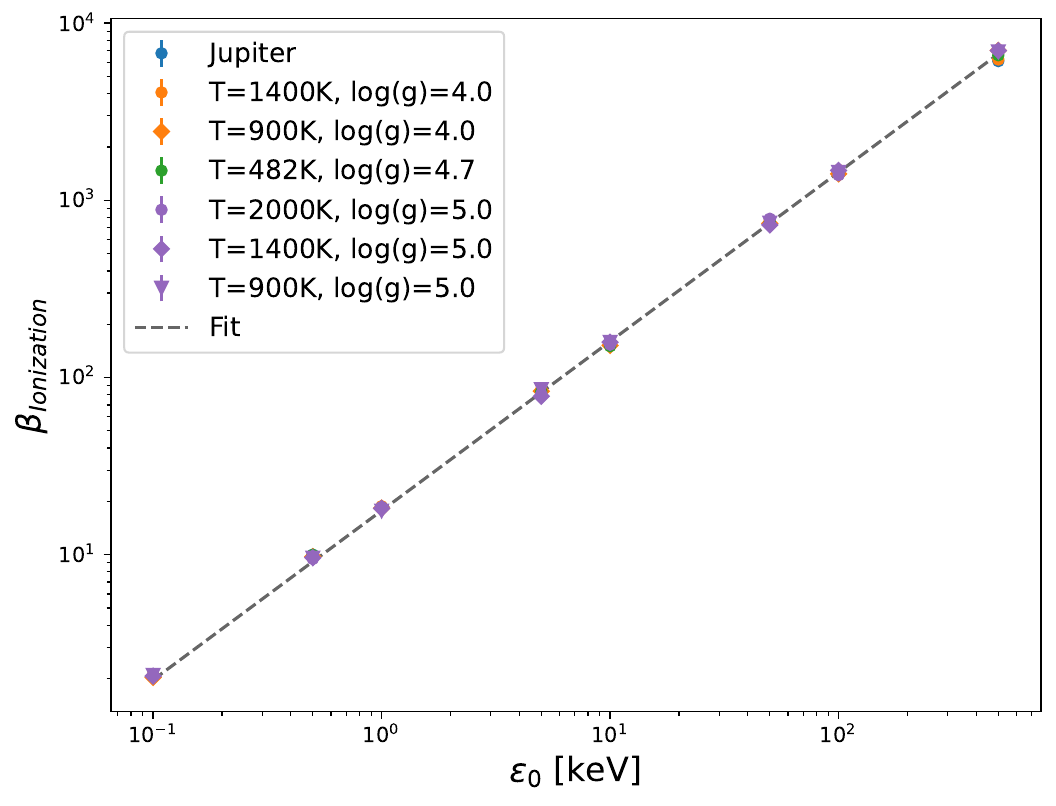}
\caption{Fit to $\beta_{ion}$, the ratio of the total number of ionizations to the number of electrons in the incident beam as a function of energy. Poisson errorbars on $\beta_{ion}$ are included but not visible. As energy increases, the number of ionizations increases because more energy is available to lose to ionizations.}
\label{fig:beta}
\end{figure}

We then need to express $q_{ion}$ in terms of the probability distribution $\mathcal{P}(N)$. First, we can recognize that $q_{ion}(z)$, the number of ionizations per length per incident electron, is the probability distribution of ionizations $\mathcal{P}(z)$ scaled by the total number of ionizations over the number of incident electrons. We can make a change of variables from $\mathcal{P}(N)$ (Equation \ref{eq:pdf}) to $\mathcal{P}(z)$. Conserving probability requires that $\mathcal{P}(z)dz = \mathcal{P}(N)dN$, or $\mathcal{P}(z) = \mathcal{P}(N)\frac{dN}{dz}$. Since column density is the integral of number density over altitude, $\frac{dN}{dz} = n(z)$. This  allows us to express $q_{ion}$ analytically using
\begin{equation}
\label{eq:qion}
    q_{ion} (z) = \beta_{ion}(\varepsilon_0) \mathcal{P}(N_{ion}(z)|\varepsilon_{0})n(z)
\end{equation}

\noindent where $\beta_{ion}$ is the unitless ratio of the total number of ionizations to the number of electrons in the incident beam, and $n(z)$ is the number density of H$_2$. The total number of ionizations increases with incident beam energy due to the availability of more energy to produce ionizations. We therefore fit a functional form to $\beta_{ion}$ (see Figure \ref{fig:beta}),

\begin{equation}
\label{eq:Nion_over_Ne}
    \ln \left( \beta_{ion} \right) = 0.955\ln(\varepsilon_0) + 2.874 
\end{equation}

We can construct similar parameterization for each type of interaction, and for the energy deposition rate. This allows us for the first time to quantify the profiles of the interactions which should produce potentially observable auroral emission for brown dwarfs. As discussed in Section \ref{sec:results}, the rates of each interaction scale with the ionization rate through the atmosphere. This allows us to simply scale the probability distribution derived in Equation \ref{eq:pdf} by $\beta_{i}/\beta_{ion}$ for each interaction $i$ (the total number of that interaction over the total number of ionizations), or equivalently to replace $\beta_{ion}$ in Equation \ref{eq:qion} with $\beta_{i}$. We show the ratios  $\beta_{i}/\beta_{ion}$ for each interaction in Table \ref{tab:interaction_scalings_vertical}. Of particular interest for studying the thermal structure of brown dwarf atmospheres is the energy deposition rate, which can be approximated as the sum of the energy deposition due to the thermalization of low energy electrons and the energy added to the atmosphere as a result of rotational and vibrational excitations of H$_2$. Note that the scaling $\beta_{E}/\beta_{ion}$ for energy deposition has units of energy because it converts number of ionizations to deposited energy, while the other scale factors are unitless.

It is important to note that the dependence of this parameterization on the properties of the object in question ($\log(g)$ and $T_{eff}$) comes only through the density profile $n(z)$, with the ionization probability density and the $\beta$ parameters depending only on incident beam energy. Because the dependence on the altitude density profile is pulled out in the fitting steps and comes in through explicit dependence of the parameterization on the density profile, the particulars of the density profiles chosen (for instance the isothermal approximation) do not effect the reported parameterization for new objects.

\subsection{Evaluation of Model and Parameterization}
\label{subsec:evaluation}
  
Next we consider the effectiveness of our parameterization by comparing it to the simulation results. We use kernel density estimation (KDE)\footnote{The \texttt{scipy} function used implements a Gaussian kernel and uses Scott's Rule for bandwidth selection} applied to bootstrapped subsamples of the histograms presented above to estimate the random errors and determine bounds on our results. 

\begin{figure}[ht]
\centering
\includegraphics[width=\columnwidth]{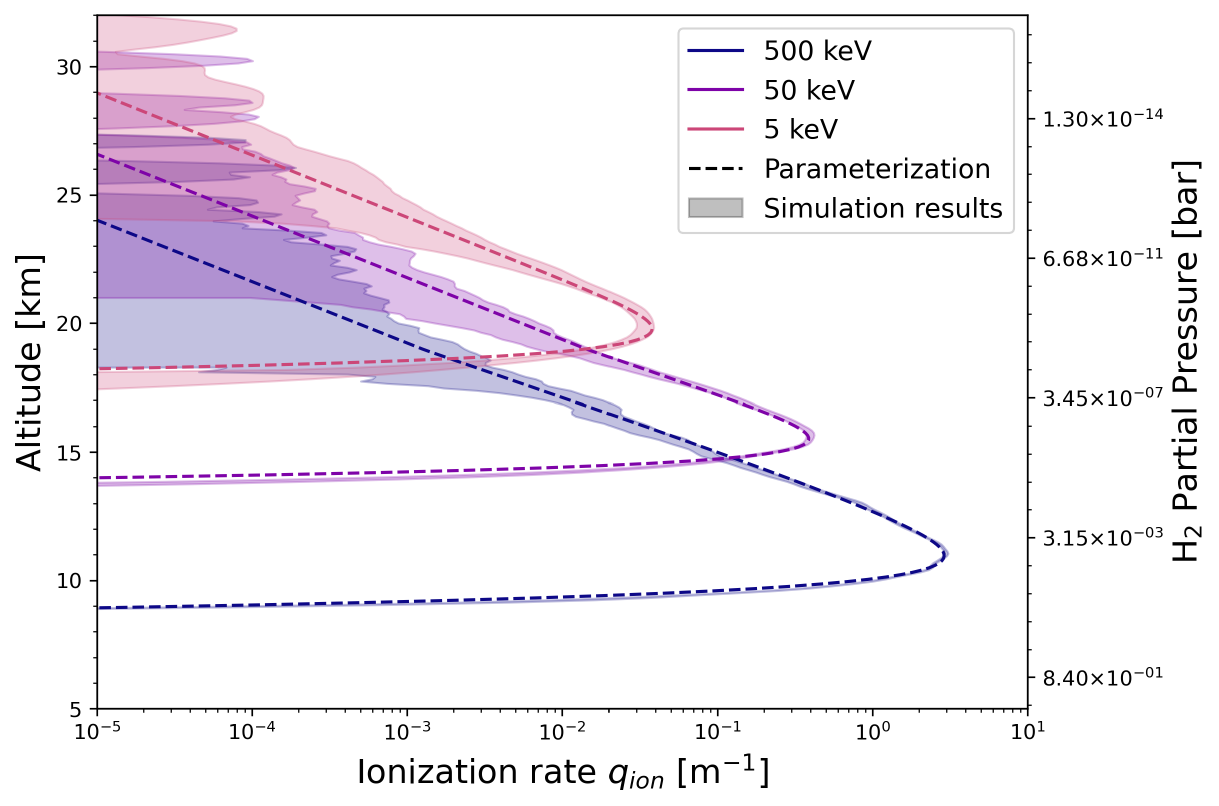}
\caption{Comparison of $q_{ion}$ from simulation results (smoothed with kernel density estimation (KDE)) to parameterization of $q_{ion}$ for the representative T-dwarf, showing the general agreement with some systematics. Bootstrapped 3$\sigma$ error region on KDE curve is shaded.}
\label{fig:900K_g5_q_kde_vs_parameterization}
\end{figure}

\begin{figure*}[ht]
\centering
\includegraphics[width=\textwidth]{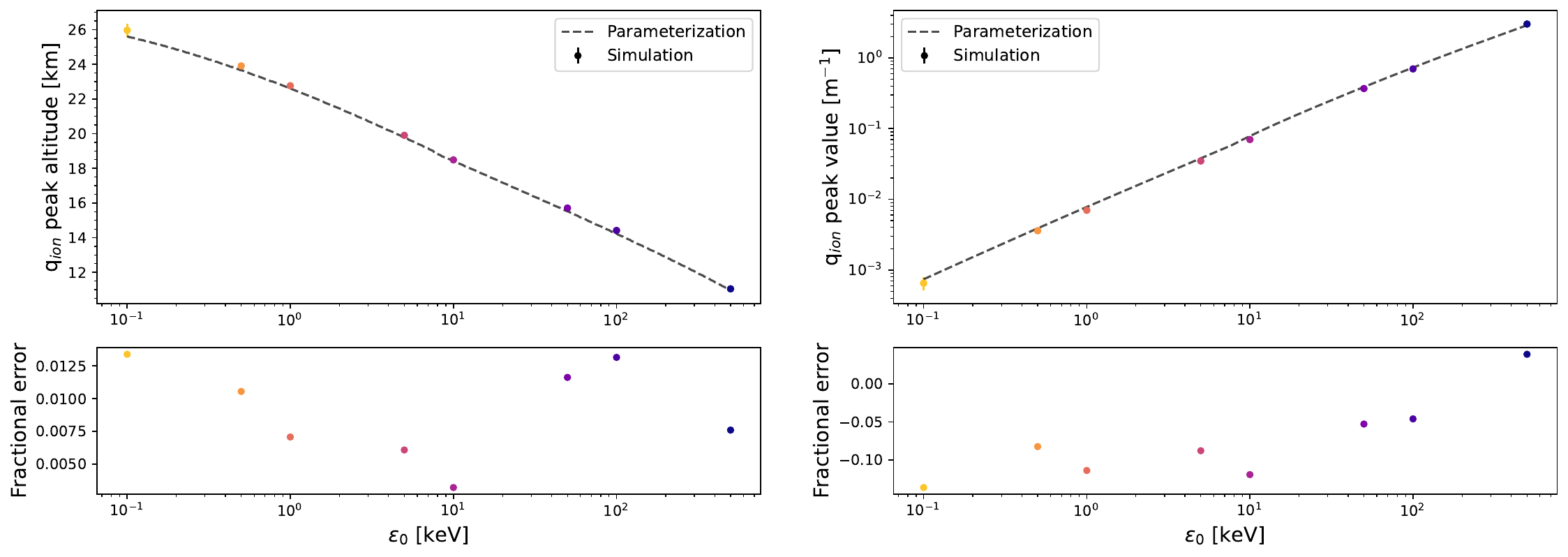}
\caption{Top panel: Ionization rate peak altitude and value for the representative T-dwarf, with 1$\sigma$ errorbars. The parameterization reproduces the simulated profiles well, with a small systematic underestimation of the peak altitude (Bottom panel).}
\label{fig:qion_residuals}
\end{figure*}

Because of the Monte Carlo nature of the simulation randomness is introduced when picking the interaction types, the energy of interaction products, and the scattering angles. Thus, each run will produce slightly different results. We can mitigate this statistical uncertainty by simply running the simulation for more electrons, and we can quantify it using a bootstrapping approach with 100 folds. For each fold we represent the distribution of interactions using KDE, and compute the values of reported parameters such as peak location and value in each fold. We report the standard deviation in parameter values between folds as errorbars in Figure \ref{fig:peak_z_and_N}. We show an example of the uncertainties in the simulation results based on bootstrapped KDE curves, with 100 folds each containing about 1500 interactions, in Figure \ref{fig:900K_g5_q_kde_vs_parameterization} for the T-dwarf. The parameterization falls into the region of uncertainty in the upper tails of the distributions, but there is a small mispecification of the peak value and location. In Figure \ref{fig:qion_residuals}  we show a comparison of the peak location and value from the KDE estimation of the simulation results and from the parameterization for $q_{ion}$ for the T-dwarf. The parameterization systematically underestimates the location of the peak by a few percent or less, and reproduces its value to within 20\%. 

\begin{figure}[ht]
\centering
\includegraphics[width=\columnwidth]{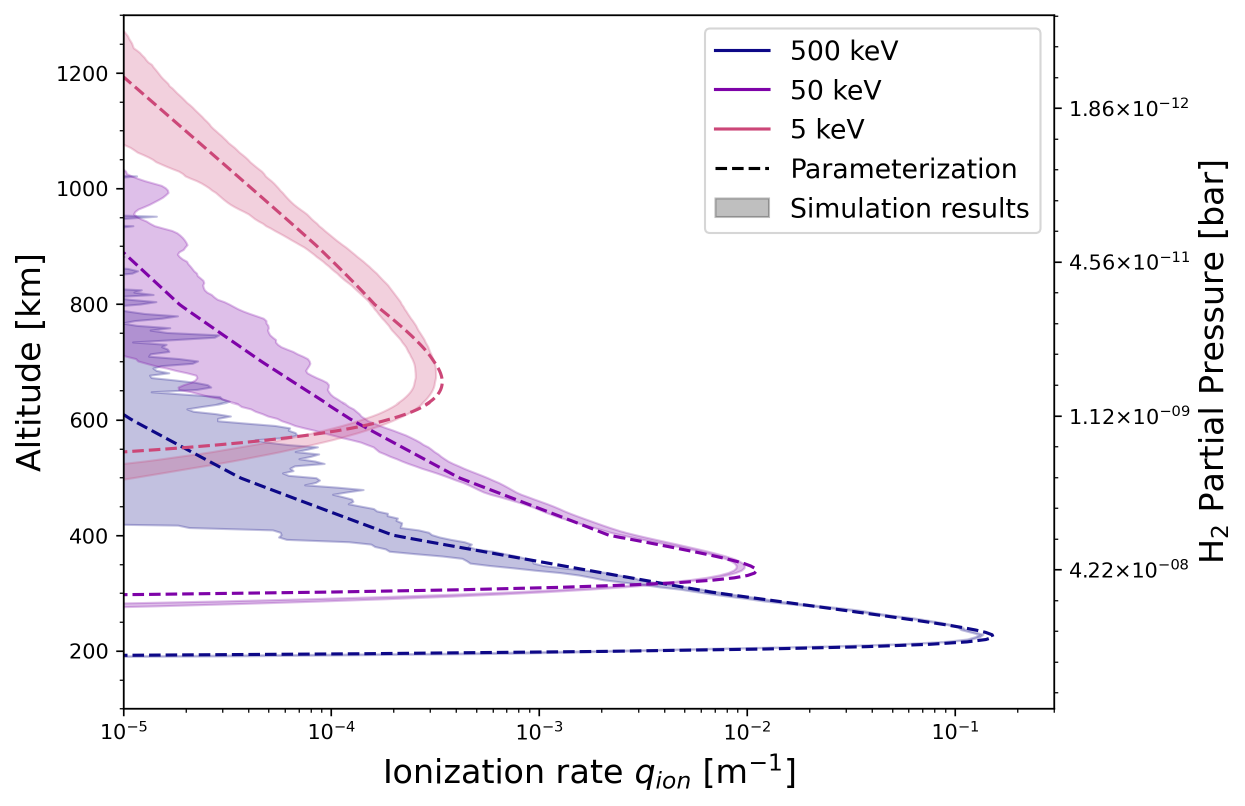}
\caption{Comparison of $q_{ion}$ from simulation results (smoothed with kernel density estimation (KDE)) to the parameterization of $q_{ion}$ for Jupiter using the Galileo derived density profile, showing the general agreement with some systematics. Bootstrapped 3$\sigma$ error region on KDE curve is shaded.}
\label{fig:Jupiter_q_kde_vs_parameterization}
\end{figure}

Our parameterization was derived from objects simulated with density profiles isothermally extended above the domain of the \texttt{PICASO} model, and so next we verify that the parameterization performs well for any density profile. We test this using the density profile for Jupiter derived from the Galileo observations. The fact that the parameterization  also matches the simulations well for the true Jupiter density profile (Figure \ref{fig:Jupiter_q_kde_vs_parameterization}) despite being derived using these isothermally extended atmospheres verifies we can successfully use this parameterization for any arbitrary density profile, not just those derived the same way as in this work.

\section{Model Applications}
\label{sec:Qion}

\begin{figure*}[ht]
\centering
\includegraphics[width=\textwidth]{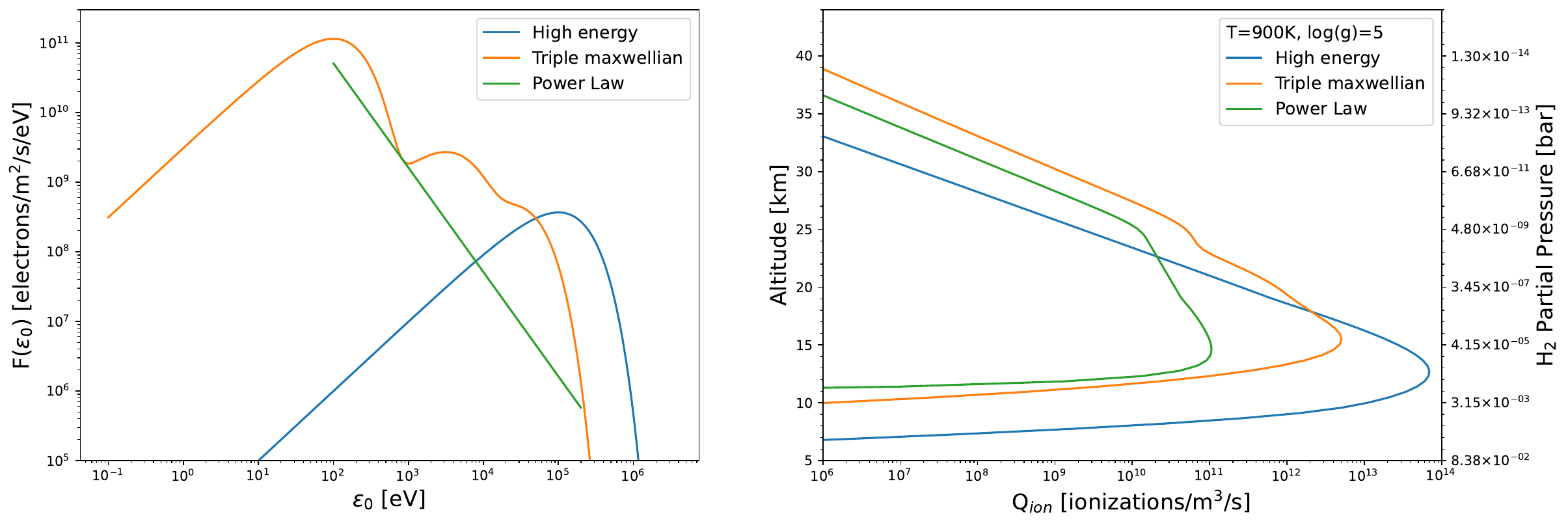}
\caption{Left panel: three representative energy spectra: a Jovian-like triple-Maxwellian spectrum, a power-law spectrum, and a high-energy spectrum, $F(\varepsilon_0)$. Right panel: $Q_{ion}$ calculated using these energy spectra and our parameterization for $q_{ion}$ }
\label{fig:Fe_Qion}
\end{figure*}

Finally, we can apply our parametrization of $q_{ion}$ to calculate the total ionization rate $Q_{ion}$ for any beam energy spectrum. This can be done by numerically carrying out the integral in Equation \ref{eq:Qion}. We can now compute $q_{ion}$ and $Q_{ion}$ for any arbitrary density profile and beam energy spectrum without running a new simulation for every possible case, purely as functions of the incident beam energy and the density profile. In Figure \ref{fig:Fe_Qion}, we show $Q_{ion}$ (right panel) for several representative electron beam energy spectra (left panel). Because the beam generation mechanism for brown dwarfs is unknown, many beam energy spectra may be possible. The triple Maxwellian spectrum is a theoretical approximation of the Jovian electron flux \citep{Grodent2001}. We compare this with a high energy spectrum to demonstrate that higher energy electrons precipitate deeper into the atmosphere, and a power law spectrum based on more recent in situ measurements of Jupiter from the Juno mission \citep{Mauk2017, Allegrini2017}. We can see that the beam energy spectra which sample higher energies produce ionization rates that peak at higher pressures and at higher values (more ionizations overall). Spectra with significant populations of low energy electrons enhance the ionization rates at higher altitudes. In this sense, the shapes of ionization rate reflects the beam energy spectrum.

The total volumetric interaction rates $Q_{i}$ for each type of interaction can be calculated using the integral in Equation \ref{eq:Qion}, once we have determined the values of $\beta_{i}$. These values are displayed in Table \ref{tab:interaction_scalings_vertical}. Of interest in particular are the volumetric even rates for B and C excitation, because these are the processes which should produce the missing auroral UV emission. The rates for all interactions, including $Q_{B}$ and $Q_{C}$, scale with $Q_{ion}$ in altitude. This means Figure \ref{fig:Fe_Qion} shows that a high energy electron beam would produce both ionizations and B and C excitations lower down in the atmosphere where greater extinction occurs along our line of sight, and where the ionization product H$_3^+$ could be more quickly destroyed by atmospheric chemistry. This supports the idea that the lack of observed UV and IR emission could be caused by a high energy beam. 

\begin{figure}[ht]
\includegraphics[width=\columnwidth]{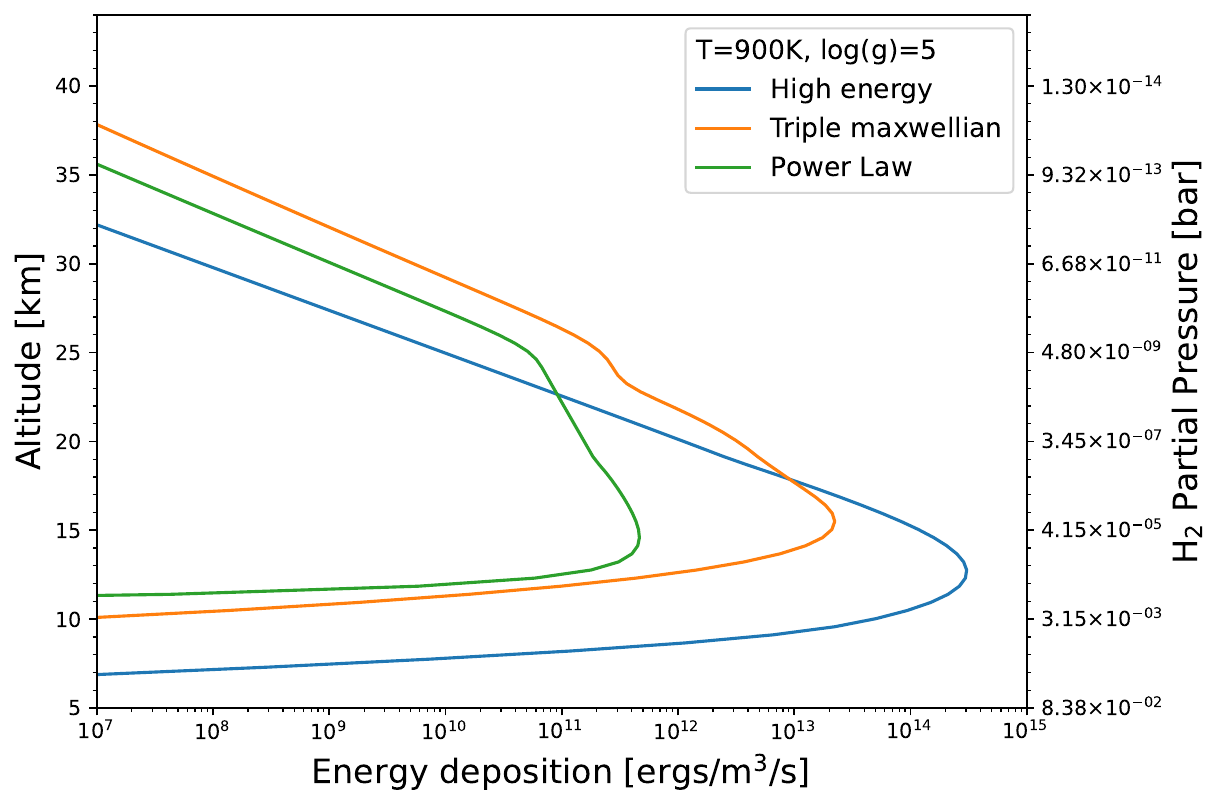}
\caption{Total volumetric energy deposition rate calculated with the example energy spectra shown in Figure \ref{fig:Fe_Qion} for the T-dwarf, using our parameterization for thermalization, rotational excitation, and vibrational excitation.}
\label{fig:Fe_Qenergy}
\end{figure}

We can similarly calculate the energy deposition rate, or heating profile. This rate is the sum of the volumetric interaction rates for thermalization, rotational excitation, and vibrational excitation, weighted by the energy deposited per interaction. The energy deposited per rotational and vibrational excitation is simply given by the excitation energies. The energy deposited per thermalized electron depends on its interaction history, and is recorded in the simulation. We show the total volumetric energy deposition rate in Figure \ref{fig:Fe_Qenergy}.

\begin{table*}[]
\begin{center}
\caption{Scalings for the rates of each interaction type, as well as minimum threshold energies for each interaction. }
\begin{tabular}{ccc}
\textbf{Interaction} & \textbf{Scaling}      & \textbf{Minimum energy {[}eV{]}} \\ \hline
Ionization           & $1$                   & $1.54 \times 10^{1}$             \\
B excitation         & $3.66 \times 10^{-1}$ & $1.12 \times 10^{1}$             \\
C excitation         & $3.4 6\times 10^{-1}$ & $1.23 \times 10^{1}$             \\
a excitation         & $1.30 \times 10^{-2}$ & $1.18 \times 10^{1}$             \\
b excitation         & $2.75 \times 10^{-1}$ & $4.48 \times 10^{0}$             \\
c excitation         & $2.10 \times 10^{-2}$ & $1.26 \times 10^{1}$             \\
e excitation         & $4.12 \times 10^{-3}$ & $1.32 \times 10^{1}$             \\
Rot. excitation      & $2.27 \times 10^{1}$  & $4.38 \times 10^{-2}$            \\
Vib. excitation      & $3.85 \times 10^{0}$ & $8.77 \times 10^{-2}$            \\
Energy deposition       & $4.39 \times 10^{0}$ eV & N/A     
\end{tabular}
\end{center}
Scalings are fits to the ratio between the curves $\beta_{i}(\varepsilon_0)$ and $\beta_{ion}(\varepsilon_0)$, where $\beta_{i}(\varepsilon_0)$ is the total number of that interaction over the total number of incident electrons. Scalings for rotational and vibrational excitation are greater than 1 due to low energy electrons which can continue scatter locally producing rotational and vibrational excitations after their energies are too low for other interactions to occur. The scaling for energy deposition provides the multiplicative factor that scales the ionization rate to produce the volumetric energy deposition rate resulting from rotational excitations, vibrational excitations, and thermalizations.
\label{tab:interaction_scalings_vertical}
\end{table*}

\cite{Pineda2024} also computed values of $Q_{ion}$, using the parameterization reported by \cite{Hiraki2008} under the assumption that it applies for brown dwarf atmospheres despite being derived using simulations of Jupiter only, and with no expectation that it should be a good approximation for brown dwarfs. We found good qualitative agreement between our calculated $Q_{ion}$ and the curves reported by \cite{Pineda2024}, ultimately justifying the assumption that the parameterization reported by \cite{Hiraki2008} extends well to other atmospheres. We observe a systematic shift of our interaction rates to lower pressures compared with the previously published results, but the effect is small. \cite{Pineda2024} concluded that the lack of observed auroral IR emission could be explained by high energy electron beams depositing energy deep into the atmosphere where ionization products may be quickly destroyed before emitting light and where extinction is high along the path to the surface. They also conclude that the width of $Q_{ion}$ in altitude or pressure is governed by the width of the beam spectrum in energy, and that $Q_{ion}$ looks very similar across objects in pressure space. Our results support these conclusions by verifying that with a more detailed simulation, including updated physics, of atmospheres across the brown dwarf parameter space and with a very different approach to parameterizing the simulation results, $Q_{ion}$ agrees with the parameterization of \cite{Hiraki2008}. By deriving a new parametrization for brown dwarfs, we quantify the results suggested by \cite{Pineda2024} and, crucially, produce profiles for each interaction important for auroral emission.

\section{Discussion}
\label{sec:discussion}
We have implemented and validated the results of a simulation of electron beam precipitation into substellar atmospheres. We have demonstrated the energy and altitude dependence of the ionization rate, as well as rates of other interactions and thermal heating. We have quantified how much deeper in the atmosphere energy deposition occurs for higher incident electron beam energies, and that the profiles for other interaction types track closely with the ionization profile. We have also shown that representing our ionization rates in reduced column density space standardizes the effects of the atmospheric pressure profile. Finally, we have developed an analytic parameterization for the interaction profile $q_{ion}$, and used it to calculate the total volumetric interaction rate and energy deposition rate.

With these results, we enable investigations into the expected auroral signatures from brown dwarfs. While interaction rates and thermal heating profiles are useful in understanding the processes driving auroral emission on brown dwarfs, future work could shed light on the expected emission produced as a result of these interactions from an energetic electron beam. It will be important to consider both the atmospheric chemistry which might produce or destroy emitting molecules, and the radiative transfer which transports photons up through the atmosphere and to our telescopes. In addition to radio ECMI emission from the electron beam itself, there are two main auroral emission processes expected based on Jovian observations: ultraviolet (UV) emission from excited atmospheric molecules, and infrared (IR) emission from ionized molecules \citep{Badman2015}.

\subsection{UV Emission}
\label{subsec:UV}

As mentioned in Section \ref{sec:intro}, UV emission is produced when atoms or molecules (in our simulation we consider H$_2$) are excited by collisions with electrons and subsequently de-excite back to the ground state, emitting a photon. In brown dwarfs this is expected based on solar system analogs to be predominantly Lyman and Werner band emission from the B and C state excitations of H$_2$, producing a forest of emission lines \citep{Gustin2013}. Despite ongoing searches for this UV emission, there has been no definitive detection yet. Our results in this work show that the B and C state excitation profiles follow the ionization profile and quantify the intensity of such excitations as a function of altitude. It is possible that the lack of observed UV emission is due to high energy incident electrons producing B and C excitations deep in the atmosphere where significant extinction occurs above the emission source. For instance, both hydrocarbons and H$_2$ molecules absorb UV emission. In situ observations of Jupiter indicate that even in the Jovian case, the beam energy spectrum could extend to MeV energies \citep{Mauk2017}. The stronger magnetic fields and fast rotation rates of brown dwarfs could produce higher energy auroral electrons \citep{Saur2021}, and work modeling the magnetospheric-ionospheric coupling of auroral currents supports higher energy electron beams \citep{Nichols2012, Turnpenney2017}. Future work could solve the equations of radiative transfer which govern the escape of this UV emission, to understand what emission might be expected under a given incident beam intensity and energy. 

\subsection{IR Emission}
\label{subsec:IR}

IR emission is produced as a result of the ionization of molecules due to collisions with energetic electrons, which can then react with neutral molecules to produce ionized products that emit in the IR. In an H$_2$ dominated atmosphere, the formation of H$_3^+$ via the reaction

\begin{equation}
    H_2 + H_2^+ \longrightarrow H_3^+ + H
\end{equation}

\noindent produces H$_3^+$, which can then emit in the IR \citep{Miller2020}. As for the UV emission discussed above, a treatment of radiative transfer through the atmosphere will allow for a better understanding of the expected emission under a specified electron beam. In addition, in the case of IR emission from ionized H$_3^+$, future work must consider another effect which can reduce the intensity of observed emission. H$_3^+$ can be destroyed through several chemical reactions, including reactions with hydrocarbons or other molecules (such as water) to form hydrocarbons and H$_2$, and recombination with electrons to form neutral H and H$_2$. The first of these processes becomes more important at higher pressures, and so if ionizations occur at high pressures in brown dwarf atmospheres, H$_3^+$ might be destroyed before it has the chance to emit in the IR, regardless of subsequent extinction \citep{Pineda2017}. 

Armed with an understanding of the expected emission intensities of different spectral lines produced by a specified electron beam energy and strength, and atmospheric profile, we can begin to assess which beam and atmospheric properties are most consistent with existing and future brown dwarf observations. An upcoming JWST program (program ID 6474) will provide new multiwavelength observations of brown dwarf aurorae, offering a basis for future comparison of simulated auroral emission with observed spectra.

In addition, a better understanding of the thermal profile can be used to provide theoretical context for recent and future observational work. For instance, as mentioned in Section \ref{sec:intro} recent JWST observations have shown an unexplained thermal inversion in the Y-dwarf W1935 \citep{Faherty2024}. JWST Cycle 4 GO proposal 7793 offers current follow-up observations of this object. This inversion requires a heating source, which could be provided by energy deposition from auroral electrons. Current work using the energy deposition rates calculated here as an energy source in modeling energy transport through the Y-dwarf atmosphere will allow us to assess this possibility. 

Understanding brown dwarf auroral beam properties could also provide insight into brown dwarf magnetospheric plasma dynamics. For  instance, if corotation breakdown plays a significant role in accelerating auroral electrons, a stronger beam will be produced for faster rotating objects with strong magnetic fields. Because of this, \cite{Saur2021} and others argue that brown dwarfs' strong magnetic fields and fast rotation rates give them the potential for much stronger auroral emission compared to gas giant planets like Jupiter. However, this mechanism requires the presence of a plasma disk, which could in theory be created via several mechanisms (in the Jupiter system, Io provides the required plasma) but which may or may not actually exist in any particular brown dwarf system. Our simulations of auroral electron precipitation could provide the basis for future work linking possible beam generation mechanisms to observables. 

\subsection{Conclusions}
In this work we present a simulation for auroral electron beam precipitation in substellar atmospheres. We demonstrate: 

\begin{enumerate}[label=(\arabic*)]
    \item Our simulation agrees well with previously published Jovian results, with differences due to updated cross sections.
    \item If the electrons incident on brown dwarf atmospheres have sufficiently high energy, this could help explain the lack of observed UV and IR emission from auroral brown dwarfs.
    \item Across objects with different effective temperatures and gravities, the distribution of interactions in column density space can be described as a function of incident beam energy alone. This means the interaction rate profile $q_{i}$ in altitude can be parameterized as a function of the incident beam energy and the atmospheric density profile. This parameterization agrees well with the simulation results.
    \item Our parameterization can be used to calculate the total volumetric rates for each interaction as well as the volumetric energy deposition rate for any specified H$_2$ dominated atmosphere and incident beam energy spectrum.
    
\end{enumerate}



\noindent This work was partially funded by Space Telescope Science Institute under award JWST 1874. This material is based upon work supported by the National Science Foundation Graduate Research Fellowship Program under Grant Nos DGE 2040434 and DGE 2137420. Any opinions, findings, and conclusions or recommendations expressed in this material are those of the authors and do not necessarily reflect the views of the National Science Foundation.


%




\bibliography{works_cited}
\bibliographystyle{aasjournal}





\appendix
\FloatBarrier

\section{Cross Sections}
\label{sec:xsecs}
\begin{figure*}[ht]
\centering
\includegraphics[width=\textwidth]{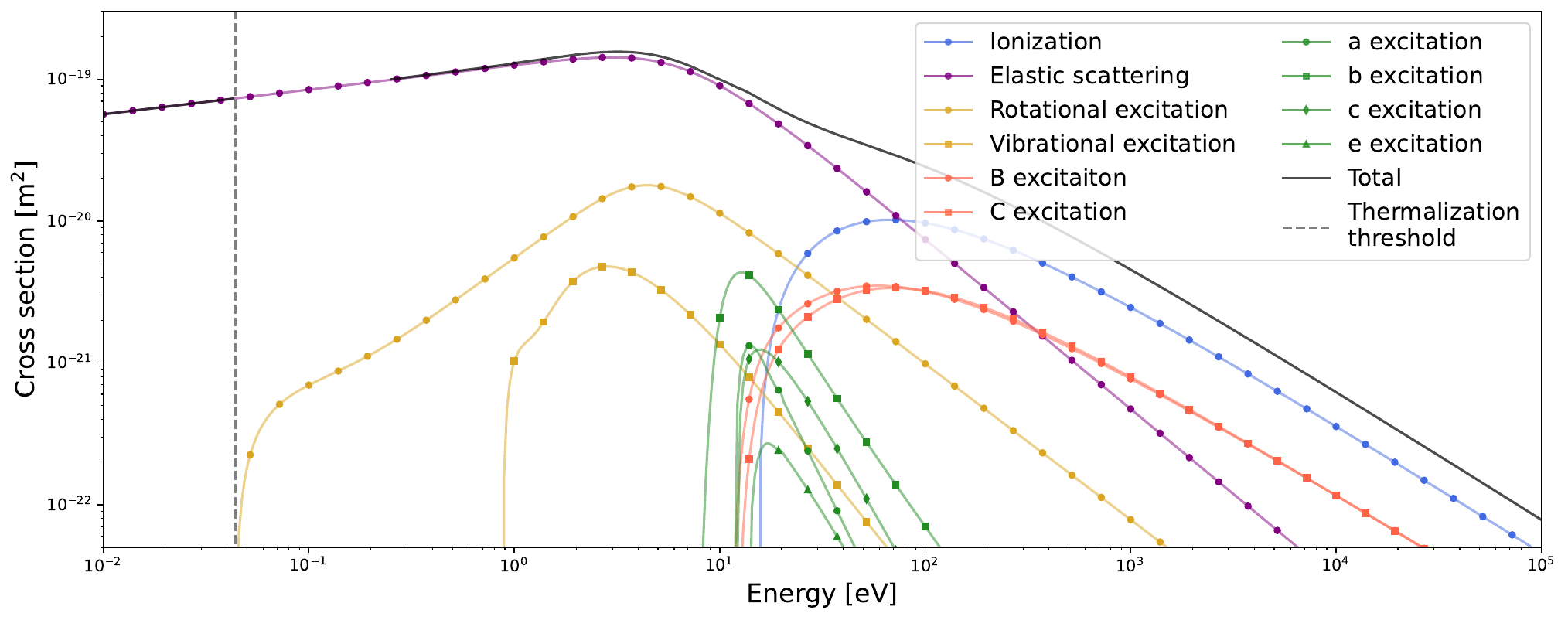}
\caption{Cross sections for the ten dominant interaction types for electrons with H$_2$, as a function of incident electron energy. At high energy ionization dominates, while at lower energies elastic scattering dominates. Various excitation processes act as energy drains across the range of incident energies. The vertical line represents the threshold energy at which electrons are removed from the simulation. The temperature used for computing the rotational cross section in this figure is that of a 900K T-dwarf, though the difference across our temperature range is small. The effect of pressure on our cross sections through modifying the initial state of H$_2$ is extremely small at the pressures and temperatures we consider, and so is neglected. For all interactions but rotational excitation (as discussed in Section \ref{subsec:params}), H$_2$ is taken to start in the ground state.}
\label{fig:xsecs}
\end{figure*}

We determine the interactions which occur at each step in the simulation using the energy dependent cross sections for each type of interaction, as shown in Figure \ref{fig:xsecs}. At high energies ionization dominates, followed by B and C excitation. At low energies, elastic scattering dominates.

\end{document}